\documentclass[preprint]{aastex}
\begin{document}

\title{Characterisation of Sloan Digital Sky Survey Stellar Photometry}

\author{Masataka Fukugita\altaffilmark{1,2,3}, Naoki Yasuda\altaffilmark{3}, Mamoru Doi\altaffilmark{4,3}, James E. Gunn\altaffilmark{5},
Donald G. York\altaffilmark{6}
}

\altaffiltext{1}{Institute for Cosmic Ray Research, University of Tokyo, Kashiwa 2778582, Japan}
\altaffiltext{2}{Institute for Advanced Study, Princeton, NJ08540, U. S. A.}
\altaffiltext{3}{Institute for the Physics and Mathematics of the Universe, University of Tokyo, Kashiwa 2778583, Japan}
\altaffiltext{4}{Institute of Astronomy, University of Tokyo, Mitaka, Tokyo 1810015, Japan}
\altaffiltext{5}{Princeton University Observatory, Princeton, NJ 08544,
U. S. A.}
\altaffiltext{6}{Department of Astronomy and Astrophysics, The University
of Chicago, Chicago, IL 60637, U. S. A.}

%--------------------------------------------------------------------------------
%--------------------------------------------------------------------------------
\begin{abstract}

   We study the photometric properties of stars in the data archive
   of the Sloan Digital Sky Survey (SDSS), the prime aim being to understand
   the photometric calibration over the entire data set.  It is
   confirmed that the photometric calibration for point sources has
   been made overall tightly against the SDSS standard stars.  We have
   also confirmed that photometric synthesis of the SDSS spectrophotometric
   data gives broad band fluxes that agree with broad band
   photometry with errors no more than 0.04 mag and little tilt along
   the wide range of colours, verifying that the response functions of
   the SDSS 2.5 m telescope system are well characterised.  We locate
   stars in the SDSS photometric system, so that stars can roughly be
   classified into spectral classes from the colour information.  We
   show how metallicity and surface gravity affect colours, and that
   stars contained in the SDSS general catalogue, plotted in colour
   space, show the distribution that matches well with what is
   anticipated from the variations of metallicity and surface gravity.
   The colour-colour plots are perfectly consistent among the three
   samples, stars in the SDSS general catalogue, SDSS standard stars
   and spectrophotometric stars of Gunn \& Stryker, especially when
   some considerations are taken into account of the differences
   (primarily metallicity) of the
   samples.  We show that the $g-r$ - inverse temperature relation
   is tight and can
   be used as a good estimator of the effective temperature of stars
   over a fairly wide range of effective temperatures.
   We also confirm that the colours of G2V stars in the SDSS
   photometric system match well with the Sun.

\end{abstract}


\section{Introduction}

The Sloan Digital Sky Survey \citep[SDSS]{York2000} provides a
large data base of photometry for galaxies and stars,
over 350 million objects in the 11663 square degree survey area,
that can be used for a great variety
of astrophysical science \citep{DR7}.  Besides
its scale, another very important feature is the special care that has been
paid to maintain the accuracy of the data, especially their photometric
accuracy, from instrumentation throughout the eight year period of the 
survey.  The photometric system is different from systems which have
been traditionally used, so it must be internally defined consistently over
that period. In addition, the survey attempted to apply photometric 
calibration to the over 1 million spectra obtained during the survey,
and we will study this calibration as well.

One of the important issues is the consistency between
spectrophotometry and broad-band photometry with the five passbands,
in the sense that the integral of spectrophotometric data with the
determined response functions should give the proper broad-band flux. This
has been a perennial problem inherent with the data bases available to
date.  Synthetic colours obtained by integrating spectrophotometric
data often give systematically increasing error with colours as
obtained by broad band photometry, typically of the order of 0.1 mag,
or sometimes larger, for stars for a range in $B-V$ from 0 to 1
\citep[see e.g.,][]{FSI1995}. This is probably largely to be attributed 
to a poor
characterisation of the response functions of conventional detector systems.

In order to minimise these systematic errors and make the broad-band
photometry as well specified as possible, much effort was invested in
the SDSS to characterise the response functions \citep{Doi2010}, in
addition to a chain of stellar calibrations \citep{Tucker2006, Stoughton2002}.
The response functions for SDSS photometry have been measured on site a
number of times during the survey, which enables us to characterise
their seasonal and secular variations. This work has been published in
\citet{Doi2010}, where the representative response functions are also
presented as the {\it Reference response functions} of the 2.5m
telescope \citep{Gunn1998,Gunn2006}. These average response
functions are recommended to be used unless the user's interests are in
fine details.  It is demonstrated in \citet{Doi2010} that the stellar
data taken in the early period of the survey are consistent with those
acquired in the later period within 0.01 mag after calibration,
including the data for the $u$ passband which has suffered from a
significant secular change in the system response and, therefore, the
expected variations in photometry have been well compensated by the
calibration process. It is concluded that the errors due to time
varying changes, both seasonal and secular, in the response
functions and calibration procedures that use several different systems
are no more than 0.01 mag as a whole.

The other special feature of the SDSS observation is hourly monitoring
of the atmospheric extinction with the ancillary 50-cm Photometric
Telescope (PT).  A set of the SDSS standard stars were observed every
night, together with hourly observation of stars in the transfer
fields from PT to the 2.5-m main telescope. The SDSS standard stars
are chosen as uniformly as possible both over the northern sky and in
colour space so that they are not inclined to give weight towards
particular classes of stars \citep{Smith2002}. For other efforts to maintain
the accuracy of the photometry, see \citet{Hogg2001,Tucker2006,Ivezic2004}.

% ubercalibration????
In this paper we study the photometric accuracies and properties of
the data presented in the SDSS catalogues, which are already in the
public domain. The purpose of this paper is to examine, in particular, 
(i) the consistency between spectrophotometry and broad-band
photometry, including the verification of the response functions for
broad band photometry, (ii) the overall accuracy of photometric
calibrations against standard stars, (iii) the location of stars in
colour space in the SDSS photometric system and its variation with the
atmospheric parameters of stars, such as effective temperature,
surface gravity and metallicity, (iv) colours of galaxies relative to
stars, and (v) brightness and colour of the Sun in the SDSS
photometric system. One of the important purposes of (iii) is to
locate stars of known spectral type in colour space so that
approximate spectral type can be inferred from colours of the SDSS
photometric system.  We also give effective temperature for those
stars. The reader is referred to \citet{Lenz1998} for an earlier study
along these lines.  There is also a substantial amount of work done
for observations of stars and for determinations of the atmospheric
parameters in {\it Sloan Extension for Galactic Understanding and
   Exploration} (SEGUE) under the SDSS project (see,
\citet{LeeI,LeeII,CAPIII,Yanny2009}). Here, the problem is that the
SDSS acquired spectra of stars fainter than $g\simeq14$, whereas the
stars for which high resolution spectra are available needed to
determine the atmospheric parameters are brighter, such as $V<14$ (NB:
$V\approx g-0.1$): hence, the overlap of the samples is small, which
makes well-calibrated stellar science with the SDSS not so straightforward. 
This
also remains as our problem and we attempt to circumvent it by the
use of bright SDSS standard stars, or of spectrophotometric standard
stars after verifying their brightness in the SDSS photometric system.

The procedure of SDSS broad band photometry is described in
\citet{Tucker2006}; see also \citet{Stoughton2002}. In brief it is
based on the network of 158 standard stars spanning a wide range of
colours, which were measured by separate observations made at USNO
\citep{Smith2002} using filters and a detector system
pseudo-equivalent to the SDSS system \citep{Doi2010} with the 
colour tansformation
between the two systems taken into account to offset the slight difference
in their response functions.
The zero point is ultimately set by the subdwarf BD$+17\arcdeg4708$ whose
brightness is defined by a synthetic calculation of the
spectrophotometric data \citep{OG1983,Oke1990} in the AB$_{95}$
magnitude system \citep{F96}.  The main telescope data are calibrated
using the set of stars which are observed simultaneously with both the
main telescope and an ancillary 50 cm Photometric Telescope, the latter
being also used to observe the SDSS standard stars whose brightness is
set by the work at USNO, and to monitor hourly atmospheric extinction
during survey operation.
\footnote{The full data release (DR7) uses
a global recalibration of the photometry using all the redundancies
in the sample, 
called \"ubercalibration\citep{Padmanabhan2008}.
We use here the standard pre-DR7 calibration;
the mean difference between photometry with the 
two calibrations is of the order of 0.001 mag and the distributions are
well described by 
Gaussian distributions with standard deviation 
0.01 mag ($g,r,i$ passband) and 0.02 mag ($u$ and $z$ passbands);
These differences are smaller than we are concerned with in this paper.}

Spectroscopy was carried out with two fibre-fed double spectrographs
covering the wavelength range 3800 to 9200\AA~ with two 2048$^2$-pixel CCD
detectors in each spectrograph. The spectrographs are fed with 640 fibers
plugged into an aluminum plate in the focal surface, positioned with the 
astrometry obtained from the imaging data
\citep{Pier2003}.  The mean
resolving power is $\Delta\lambda/\lambda\simeq 1850-2200$ depending on the
wavelength.
Spectrophotometric calibration
is carried out using spectrophotometric standard stars selected in
each field. Standard stars were intended to be typically F8 dwarfs \citep{DR2}
but the set in actuality
contains a variety of similar spectral types
\citep{DR6}; 
the spectra are fit to models given by Gray et al (2001), and
the model spectra are used for calibration.
Spectrophotometric calibration also
requires match between spectrophotometric fluxes and the
broad-band photometric PSF fluxes. This is done
with a single calibration determined for each camera in a given 
plate \citep{DR6}.

The data acquired in the SDSS projects are all published in 7 data
releases, DR1$-$DR7 \citep{DR1,DR2,DR3,DR4,DR5,DR6,DR7}. All DR data
are cumulative, and DR7 stands for the final data release of the SDSS
I and II, which employ the original instrumentation of the SDSS. 
We use the data set from DR6 in the study presented in this
paper. The photometric calibration is made against the set of standard
stars.  Reddening corrections are not applied for the stars.  The
stars are located mostly in low extinction regions.  If we take the
extinction map of \citet{SFD1998}, the peak of the distribution is at
$(B-V)\approx 0.02$ with the mean 0.04 and the variance 0.03; only
2.6\% of stars are located in the region $E(B-V)>0.1$.  Note that this
reddening is the maximum value along the line of sight and the actual
reddening for objects within the disk of the galaxy is smaller.

\section{Comparison between spectrophotometry and broad-band photometry}

In this paper we consider synthetic fluxes, obtained by integrating 
with the $d\ln \nu$ measure over the
spectrophotometric data with the response functions for each filter.
The flux calibration of the
spectrophotometry was done by comparing the spectrophotometric data with
the PSF fluxes of the standard stars in each spectroscopic field (actually
two fields independently, one for each spectrograph.)
% from the time of DR6 is made against the PSF flux of
%a {\it single} star chosen in each camera field. 
So, it is important
to confirm that synthetic spectrophotometric fluxes indeed agree with broad
band photometric fluxes within tolerable errors for the majority of
stars across the wide spectral range.

\begin{figure}
\plotone{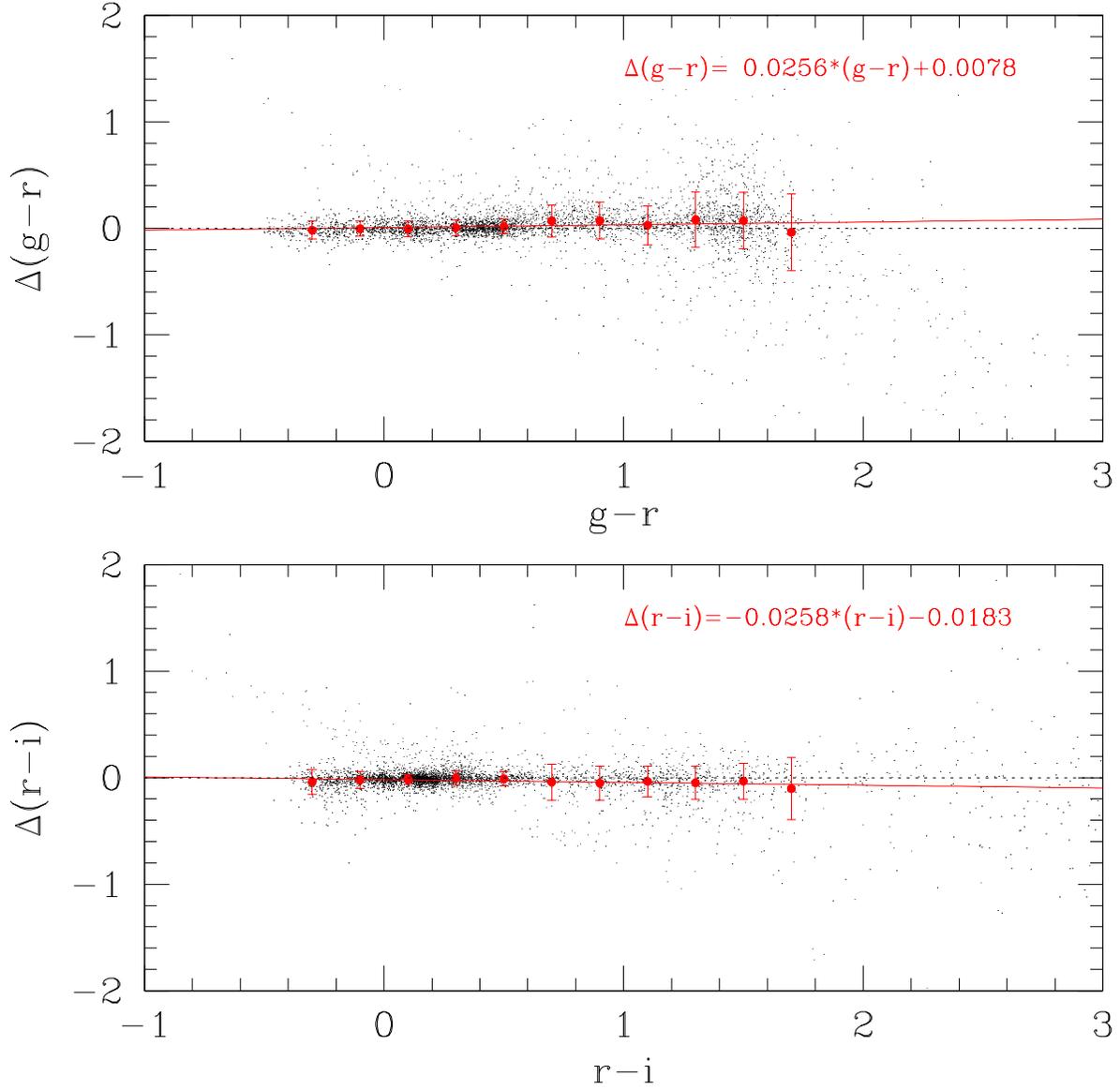}
\caption{Difference in the brightness of broad band colour between
the synthesis of the spectrophotometric data and the direct 
broad band photometry,  $m({\rm spectro})-m({\rm photo})$, is plotted 
against colours of stars
acquired in the southern equatorial stripe (Stripe 82) for $g-r$ and
$r-i$. The lines are linear fits with the coefficients given in the
figure.
\label{fig:1}}
\end{figure}

In Figure \ref{fig:1} we present the difference of synthetic colours
and photometric colours for (a) $g-r$ and (b) $r-i$ as a function of
colours of stars. Stars are taken from the observation for southern
equatorial stripe (stripe 82), and the psf magnitude is adopted for
broad band photometry. The error bar shows the variance in each colour
bin, 0.2 mag in $g-r$ or $r-i$. 
The offset of the mean
is $<0.05-0.07$ mag for most of the colour bins, and it increases to
0.08 mag for red stars with $g-r\approx 1.3$.  The rms scatter we have
seen is about 0.1 mag, and it goes up to $>0.15$ mag towards redder
stars with $g-r>0.7$.  This is compared with the rms error quoted
for the spectrophotometric calibration, 0.04 mag (DR7). The smaller scatter 
for bluer stars offsets the larger rms scatter in
longer wavelengths for the total sample.

The $u$ and $z$ passbands are out of 
the range of our
spectroscopy, which spans from 3800\AA~to 9200\AA~ and therefore are
not shown.  We adopt the SDSS spectroscopic data and
use the reference response function for the SDSS 2.5 m telescope
presented in \citet{Doi2010} referred to the standard 1.3 airmass
value for the atmospheric extinction corrections.

The global tilt of the offset across colour
is small.
After 3$\sigma$ rejection of highly deviant data points, the
mean gives $\Delta(g-r)_{\rm spec-photo}=0.026(g-r)+0.008$ and
$\Delta(r-i)_{\rm spec-photo}=0.026(r-i)-0.018$ for the range $-0.4 <
g-r < 1.8$ and $-0.4 < r-i < 1.8$, respectively. 
This implies that the response curves are well characterised.
Synthetic colours may be substituted for broad band measurements at
least in the mean over the stars in the sample and for non-variable
stars.  We consider that this is an improvement, crudely speaking, of
a factor of 5 in accuracy compared with previous `best'
spectrophotometry-photometry data sets \citep[e.g.,][]{F96}. Highly
deviant data points are likely to be variable stars or double stars,
as found for subsamples \citep{Sesar2007}.

\section{Colours of stars in the SDSS data base}

\begin{figure}
\plotone{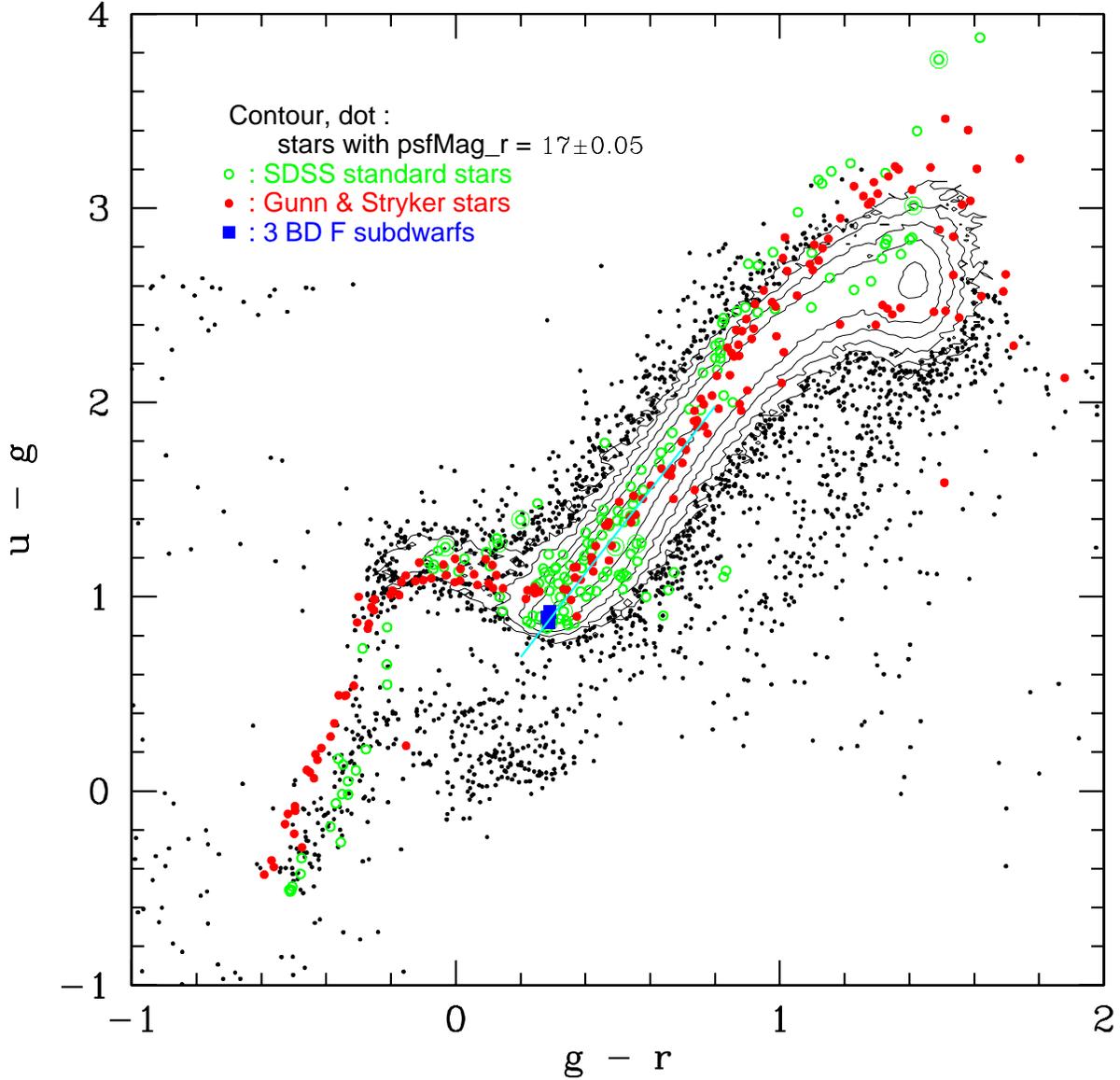}
\caption{Three hundred thousand stars in $u-g$ vs $g-r$ colour space
taken from SDSS general star catalogue. Densely populated regions 
are shown by contours, and one step corresponds to $\sqrt{10}$ increase
in the density
with the outermost contour for 75 stars/(0.1 mag)$^2$.
Open circles show SDSS standard stars, and solid circles are the GS83 stars.
Those denoted by solid squares are three BD F-subdwarfs, taken to
give the observational flux standard. The oblique 
line segment along the contour
(from $g-r=0.2$ to 0.8) indicates the approximate positions of 
the centre of gravity 
for the star distribution used to define $d_{ug}$ used in later
figures.
\label{fig:2}}
\end{figure}

In Figure \ref{fig:2} we plot $u-g$ of 290647 stars with
$16.95<r<17.05$ against $g-r$ taken from the full DR6 catalogue.
Regions of densely populated points are represented by contours, which
correspond to an increase of density by $\sqrt{10}$ in each step.
Those denoted by individual points are 2660 stars, or 0.9\% of all
stars.  The catalogue contains an additional 583 stars that fall
outside the range and are not plotted.  We also superimpose the data
of 158 SDSS standard stars given in \citet{Smith2002} and 175
spectrophotometric stars given by \citet[hereafter GS83]{GS}. The
colour of GS83 stars is derived by synthetic calculations with the
reference response function of the 2.5m telescope system at 1.3
airmass after taking the redleak of the $u$-filter into
account\footnote{\citet{Lenz1998} have also given SDSS colours for the
GS83 stars using synthetic calculations. Their colours, however, do
not agree well with ours, especially towards red stars (the difference
becomes as much as 0.5 mag for $u-g$ colour with $g-r\approx 1$
stars).  This is much more than we expect from the difference between
the two response functions, \citet{F96}, which was adopted by Lenz et
al., and \citet{Doi2010} adopted here.  We expect that the two lead to
a difference generally smaller than 0.02 mag and, at largest, $<0.04$
mag for $u-g$ colour of red stars.  We could not find the origin of
this gross discrepancy.}.
Since the GS stars are mostly located in the low latitude field and
receive large reddening, dereddening correction is applied
as in GS93. It is noted that the reddening model used in
that paper is probably oversimple, although the resulting colour
does not differ too significantly from what is calculated
using a more standard extinction curve.

There are 5 GS83 stars that fall beyond the figure frames
and are dropped. All SDSS standard stars are shown.  The redleak
affects $u-g$ colour by no more than 0.02 mag for stars with $g-r\le
1$, while it increases to 0.2 mag close to the red edge of the contour,
$g-r\approx 1.5$.  Recognition of this correction improves agreement
between the GS83 stars and other stars, if only slightly.  The three
solid squares around $g-r\approx 0.28$ represent the metal poor F
subdwarfs, BD$+17\arcdeg4708$, BD$+26\arcdeg2606$ and
BD$+21\arcdeg0607$, the latter two of which were added to the first to
set up the network of F-subdwarf standards approximately
8 hours apart \citep{Smith2002}.  Accurate spectrophotometry is
available only for the first two stars \citep{OG1983, Oke1990}, but
the third is sufficiently similar to the first two that the adopted
linear interpolation to obtain its fluxes should be adequate.

We confirmed that the contours of the distribution of stars 
change only moderately when we take stars of $r\simeq 18$ or 19 instead of
$r\simeq17$.
The distributions change in detail, doubtless due to the changes in stellar
populations from old disk to thick disk and halo as one goes fainter
\citep{BChen2001,Ivezic2008}.
The contours, however, overlap
very well for stars bluer than $g-r\simeq 1$ for the 3 samples that differ
in brightness, indicating that the calibration is robust against
brightness of stars.

We see that both SDSS standard stars and GS93 stars are located near the
centre of gravity of the distribution of stars in the SDSS general sample
for $g-r<0.8$. This also serves, albeit indirectly, as the evidence that
spectrophotometry of the GS83 stars and 
spectroscopic synthesis give correct broad band colours to the
level we have required.
We see for red stars with $g-r>0.7$ some shift of the loci of these stars 
towards the upper side,
$u-g$ being redder by  $\approx0.3$ mag. On the other hand
there appears a second locus of the GS83 stars, which is shifted {\it
downwards} by $\approx 0.1$ mag with respect to the centre of gravity. We
ascribe this to widening of the gap in colour space between the distribution of
luminosity class III and V stars for $g-r\geq 1$, as shown in the next
section.  For red stars we observe that the locus of
the GS83 stars is more heavily populated around the upper side, to 
which stars are driven by
high metallicity (nearly solar) of the GS83 stars.
In general, the scatter becomes larger towards the redder end, as a 
natural wilder variation of spectra due to molecular
bands developed in later K to M stars for $g-r\ge 1$.  The span of the
$u-g$ colour
range against the variation of metallicity, and also of gravity,
increases towards the reddest stars.

% ?????? Masataka, read this
For very blue stars with $g-r<-0.2$, the locus of the GS83 stars does
not agree with that of SDSS standard stars and that of the general
SDSS stars, by 0.1 mag in $g-r$ or as much as 0.4 mag in $u-g$.  This
is significantly larger than could arise from the different
response functions of SDSS specifications.  One might suspect that this
displacement is due to reddening, especially given the crudity of the
GS reddening corrections. In fact, however, the displacement corresponds
approximately to $E(B-V)\approx 0.15$.  Such large extinction
is very rare in the SDSS data, but is present in essentially all the
blue GS stars, and is poorly determined.
However, if such corrections are systematic and are made, we would
expect shifts in $i-z$ vs $g-r$, for which the shift
(see Figure \ref{fig:6} below), if any, takes place in a way opposite
to that expected.  Note that the standard extinction curve \citep{ODonnell}
gives reddening for $r-i$ and $i-z$ smaller than that for $u-g$ only
by a factor of two (see arrows for extinction in Figure
\ref{fig:8}-\ref{fig:10} below).  In $r-i$ vs $g-r$ the two loci
overlap at a good precision; see Figure \ref{fig:5} below, but the
extinction for this colour is nearly parallel to the stellar locus,
and it is not easy to separate out extinction.  We thus do not find an
easy explanation that accounts for the shift in $u-g$ vs $g-r$ plane between
the GS83 stars relative to the SDSS stars.  We do note, however, that the
{\it populations} of the blue GS stars and the blue SDSS stars are 
completely different; the blue GS stars are all main-sequence disk Population
I objects, and the blue SDSS stars are essentially all low-mass, hot, 
highly evolved Population II objects of a variety of types, so the 
discrepancy may not be surprising.
We note also that the locus of
the SDSS stars overlaps well that of the SDSS standard stars.

\begin{figure}
\plotone{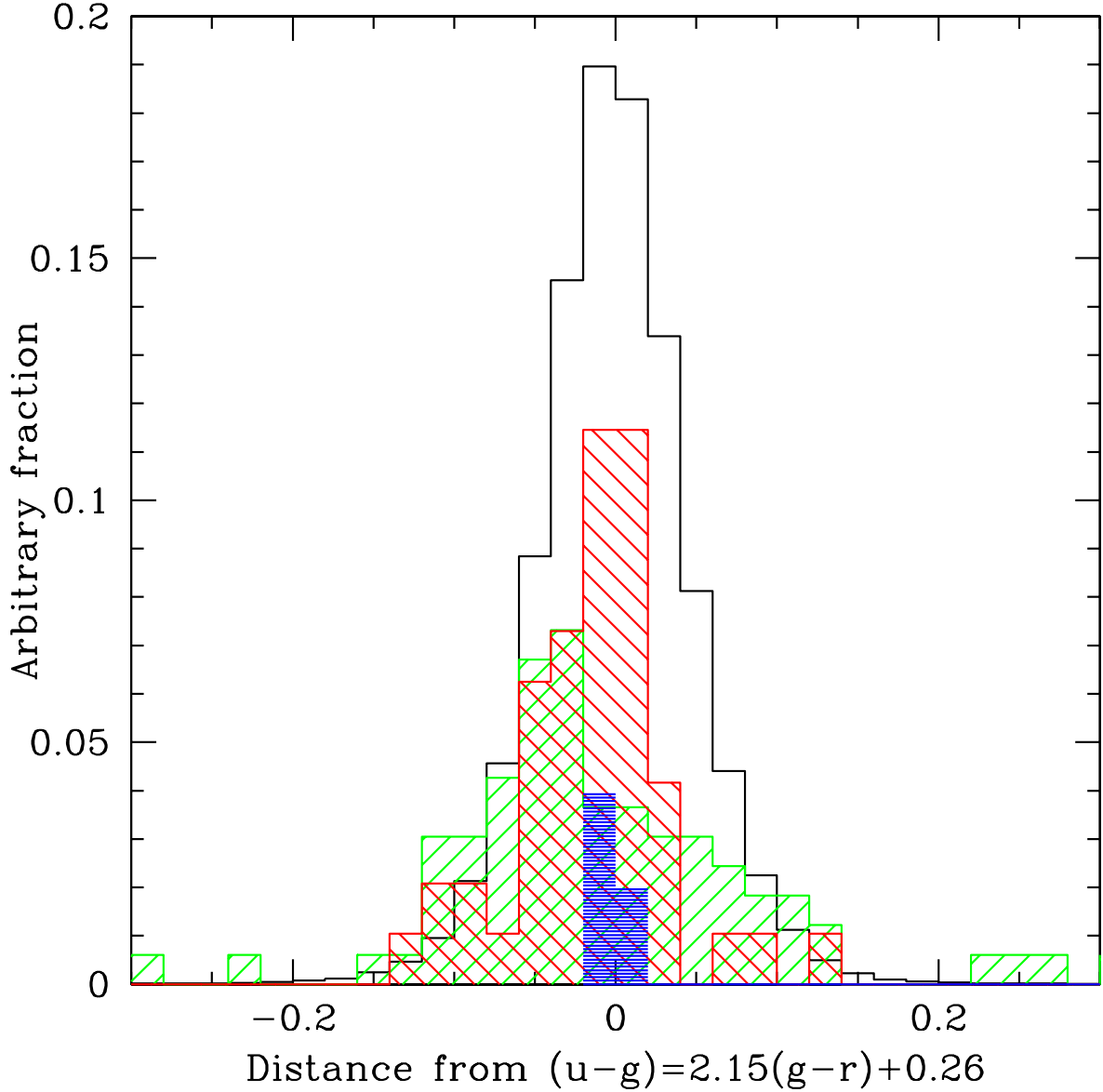}
\caption{Distribution of stars with $0.2<g-r<0.8$ in different samples
projected to the axis that is perpendicular to the distribution of
densely populated region, which is around $u-g=2.15(g-r)+0.26$
in $g-r$ vs $u-g$ colour space, shown in Figure 2. %\ref{figure2.ps}.
The abscissa is $d_{ug}$ (mag).
Open histogram shows the stars in the general SDSS catalogue, 
the histogram hatched
from up-left to down-right (red) shows stars in the GS83 sample, and that
hatched from up-right to down-left (green) shows the SDSS standard
stars presented in \citet{Smith2002}.  The thick horizontally hatched
histogram at around the zero point in the abscissa 
shows 3 F subdwarfs taken as the fundamental standard at
the observation.  The ordinate is in arbitrary units.
\label{fig:3}}
\end{figure}

To scrutinize the possible offset among the populations in three
different star samples, i.e., the sample in the SDSS general
catalogue, the SDSS standard star sample and the GS83 sample, we plot
in Figure \ref{fig:3} the distribution of the three populations,
projected onto the line orthogonal to $(u-g)=2.15(g-r)+0.26$
which follows the locus of the stellar colour for $0.2\leq g-r\leq
0.8$, indicated as the line segment in Figure \ref{fig:2}. The
distribution of the three F subdwarfs is also shown.  The distance
along this base line measured from the locus of the population is
denoted by $d_{ug}$ (mag) with the positive direction towards the lower
right (increasing $g-r$ and decreasing $u-g$).  We see that all
populations are distributed very well with respect to the common zero
point, although the distribution for the SDSS standard stars
is somewhat skewed towards the bluer side, which
can be understood by a larger weight of lower metallicity stars in that sample.
No alarming offset is visible in the zero point among the four samples beyond
$\approx 0.05$ mag in $d_{ug}$.

\begin{figure}
\plotone{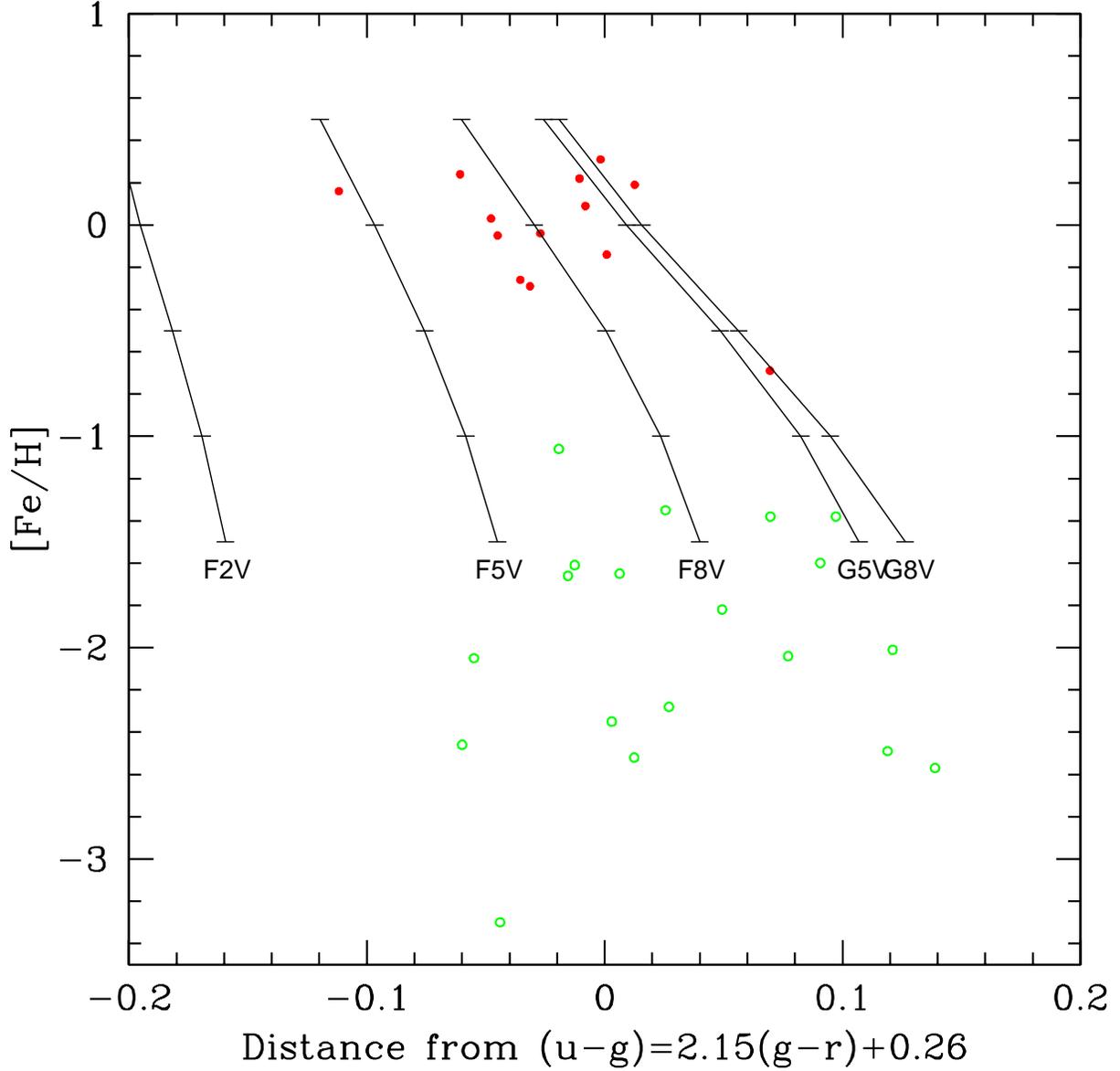}
\caption{Correlation of [Fe/H] estimated in \citet{CdS2001}
with $d_{ug}$, i.e., 
the distance from the line $(u-g)=2.15(g-r)+0.26$,
which represents the centre of gravity of the 
most densely populated region between $g-r=0.2$ and
0.8, as
  defined in the text. 
The sample is taken from the GS83 stars (solid circles) and
the SDSS standard stars (open circles).
The line segments are the metallicity dependence predicted by
the Kurucz atmosphere model \citep{CK2004} from F2 to G9 stars 
in the main sequence.
\label{fig:4}}
\end{figure}

The distance $d_{ug}$ is correlated with metallicity.  21
out of the 158 SDSS standard stars and 44 out of the 175 GS83 stars are
given estimates for metallicity with high resolution spectroscopic
observations \citep{CdS2001}.  Figure \ref{fig:4}
shows those stars having colours in the range $0.2<g-r<0.8$ plotted in the
plane of metallicity versus $d_{ug}$, indicating the correlation of
metallicity with  $d_{ug}$. It shows 
that the GS
stars plotted in this figure have metallicity close to solar, while
many of the SDSS standard stars are indeed metal poor [Fe/H]$<-1$, as
expected from the fact that they are mostly high latitude halo objects
\citep{Ivezic2008}.
Although these are partial samples of the GS83 and the SDSS standard
stars, this explains the difference of the two samples we saw in the
$u-g$ vs.  $g-r$ plane.  We also plot the bundle of line segments
which show the prediction of the Kurucz atmosphere model \citep{CK2004}
for stars with F2V to G8V, metallicity from [Fe/H]=0.5 being at the top
to $-1.5$ at the bottom.
% \footnote{In the range that concerns us here
%\citet{Kurucz1993} and \citet{CK2004} give similar results.}.

A similar plot is shown in Figure \ref{fig:5} for $r-i$ vs. $g-r$.
The distribution is narrower and the agreement among the three
distributions is much tighter than we have seen for $u-g$ vs.  $g-r$.
The SDSS standard stars are located very close to the centre of
gravity of the general star sample. The GS83 stars are also located
close to the centre of gravity: we observe a slight displacement
between the SDSS standard star sample and the GS83 star sample, which
is of the order of the difference we expect from the different
metallicity.  The three F subdwarfs are also on the top of the loci.
The difference in the loci between the luminosity class III and V is
negligibly small, as we see in the next section more quantitatively.

A similar observation holds for the plot of $i-z$ against $g-r$ shown
in Figure \ref{fig:6}.  We do not observe displacements among the
three populations, again except for some disagreement for
very blue stars. Otherwise, 
the distributions of the standard stars and GS83 stars
agree well with those of the stars in the  SDSS general catalogue.

These considerations mean that the stellar loci are well defined in
the SDSS photometry and depend little on the sample taken, and also
verify that the photometric calibration has been made well in
colour against the standard stars. This does not
mean that the broad band flux is accurate in physical units (AB
magnitude system), however.

\section{Variations of colours and stellar atmosphere}

The main parameters that regulate colours of stars are effective
temperature $T_{\rm eff}$, surface gravity $g$ and metallicity [Fe/H].
We first show that $T_{\rm eff}$ is well represented by $g-r$ colour
(see also \citet{Ivezic2008}).
The most accurate measurement of $T_{\rm eff}$ is 
made with Michelson
interferometry of stars to measure their diameters.  This has been done,
however, only for very bright stars and the sample is very limited.
%none of them overlap with the samples that are endowed with SDSS photometry. 
The sample may be extended by stars for which
temperatures have been obtained from the infrared flux method (IRFM) 
\citep{BS1977},
which is calibrated against stars with Michelson interferometry 
measurements of high accuracy.  The IRFM
suffers from less model dependence; temperatures from the IRFM method
%has been calibrated against the Michelson interferometry measurement
%with high accuracy: the accuracy of the temperature 
are supposed to be accurate at the
1\% level.  The stars with effective temperatures estimated by IRFM are
still too bright for the main sample of SDSS, but the GS83 sample 
has stars whose
effective temperature has been estimated with IRFM.  The argument
in the previous sections ensures that photometry from spectrophotometric
synthesis for
GS83 stars is sufficiently accurate.

\begin{figure}
\plotone{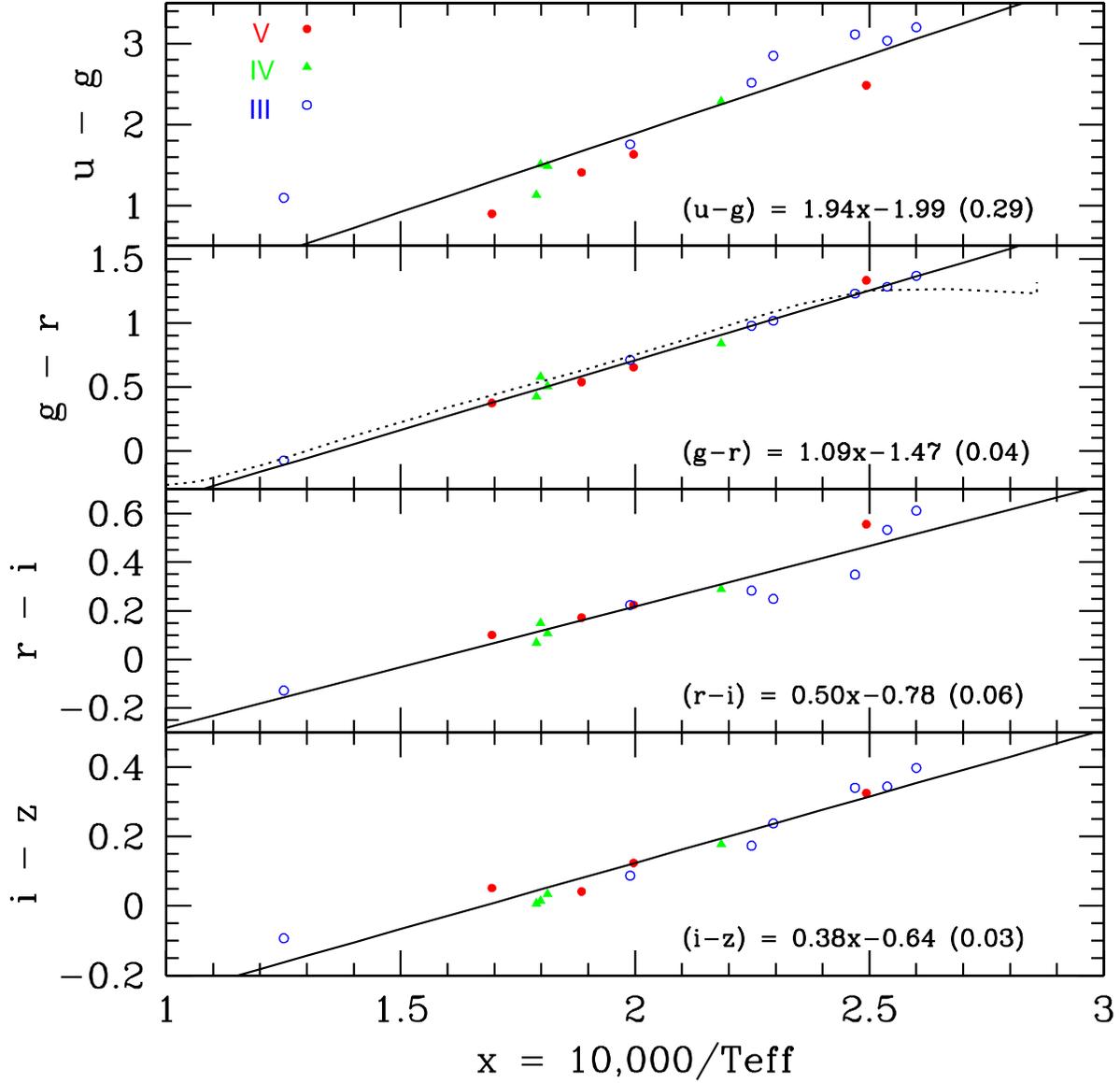}
\caption{$u-g$, $g-r$, $r-i$ and $i-z$ colours plotted against inverse
temperature in units of 10000K. The GS83 stars for which temperature is
estimated using the IRFM are shown.  The solid circles (red) show stars
with the luminosity class V, open circles (blue) stars with luminosity
class III and solid triangles (green) stars with luminosity class IV.
Lines are linear fits to the data and their coefficients given in 
the figure where parentheses
are dispersion of the fit in colour units. In the fit 
$x=10^4{\rm K}/T_{\rm eff}$
and $y$ is the corresponding colour.
The dotted curve in the panel for $g-r$ is the prediction of the
Castelli-Kurucz atmosphere model for main sequence stars. 
\label{fig:7}}
\end{figure}

Figure \ref{fig:7} shows colours of 15 GS83 stars against inverse effective
temperature that is given in the IRFM
catalogue of \citet{DiBenedetto1998}.  Among the panels that show
$u-g$, $g-r$, $r-i$ and $i-z$ colours the correlation with $g-r$
colour is tightest: the stars are well fitted with the simple formula
\begin{equation} 
T_{\rm eff}/10^4{\rm K}={1.09 \over g-r+1.47}
\label{eqn:1}
\end{equation}
with an rms scatter around the fit $\sigma_{g-r}=0.04$ or
$\sigma(T_{\rm eff})=93$K.  Note that the star sample contains
different luminosity classes, III (7 stars), IV (4 stars) and V (4
stars), and metallicity is not differentiated.  We do not observe any
systematic difference for $g-r$ among the different luminosity
classes. We note, however, that this colour-temperature relation is empirically
verified only in the range, $1.25<10^4{\rm K}/T_{\rm eff}<2.6$ or
$3850{\rm K}<T_{\rm eff}<8000{\rm K}$. The use of the Castelli-Kurucz
atmosphere model (indicated by the dotted curve in the second panel of
Figure \ref{fig:7}) suggests that this inverse linear relation may break down
for $T_{\rm eff}>9500{\rm K}$: for those stars with temperature higher
than $T_{\rm eff}\approx 8500{\rm K}$, $T_{\rm eff}$ is higher than
eq.(\ref{eqn:1}) indicates for given $g-r$.  A departure from
eq.(\ref{eqn:1}) is also indicated towards lower temperature below
3850K.

The expression we derived above agrees well with the $\log T_{\rm
eff}-(g-r)$ colour relation given by \citet{Ivezic2008} but only in a
finite segment, $-0.3<g-r<0.7$, or $5000<T_{\rm eff}<8000$K.  The
latter relation gives temperature significantly lower than the IRFM
estimate for stars redder than $g-r\simeq 0.7$.  \citet{LeeI} use a
polynomial which is third order in $g-r$ to estimate $T_{\rm
eff}$ from $g-r$ in the SEGUE stellar parameter pipeline.  It agrees with
eq. (\ref{eqn:1}) within 0.02 dex for the range $-0.6<g-r<1.2$
($4200K<T_{\rm eff}<12500$K).

We see some systematic shifts among different luminosity classes for the
$u-g$ vs. inverse temperature plot; e.g., all 4 class V stars lie below the
fit line, and 6 of the 7 class III stars lie above the line.  The $i-z$
colour is also well correlated with temperature, the dispersion of
$i-z$ being also small (0.03).  The corresponding dispersion of temperature,
$\sigma(T_{\rm eff})=190$K, however, is twice as large as that for the
$g-r$ inverse temperature relation due to the shallower slope.

Metallicity has been estimated with high resolution
spectroscopy \citep{CdS2001} for 9 among the 15 stars 
given in
this figure and it ranges from [Fe/H]=$-$0.69 to 0.31.  We do not
observe a systematic trend with metallicity in the
$g-r$ vs. $T_{\rm eff}$ plot. The variation with metallicity is
significantly smaller than is seen with $B-V$, since the metal lines
are fewer and weaker in the $g$ (and $r$) passband than in the
$B$ passband.
%this feature was built into the design of the SDSS filters in the beginning. 
The scatter around the colour temperature relation is
nearly as small as the one that would be obtained after the
metallicity effect taken into account in the $B-V$ - $T_{\rm eff}$
relation, or the $V-K$ - $T_{\rm eff}$ relation for which the smallest
scatter has been claimed \citep{Cohen1978, DiBenedetto1998}.
Some metallicity trends are observed for the $u-g$ - $T_{\rm eff}$ plot
in the way one would expect, but the present sample is too scanty to
derive the metallicity dependence in a meaningful way. 
We conclude that $g-r$ colour is a
good indicator for (inverse) temperature, much better than $B-V$ because of
weak metal lines in the $g$ and $r$ passbands.

\begin{figure}
%\plottwo{figure8a.ps}{figure8b.ps}
\plotone{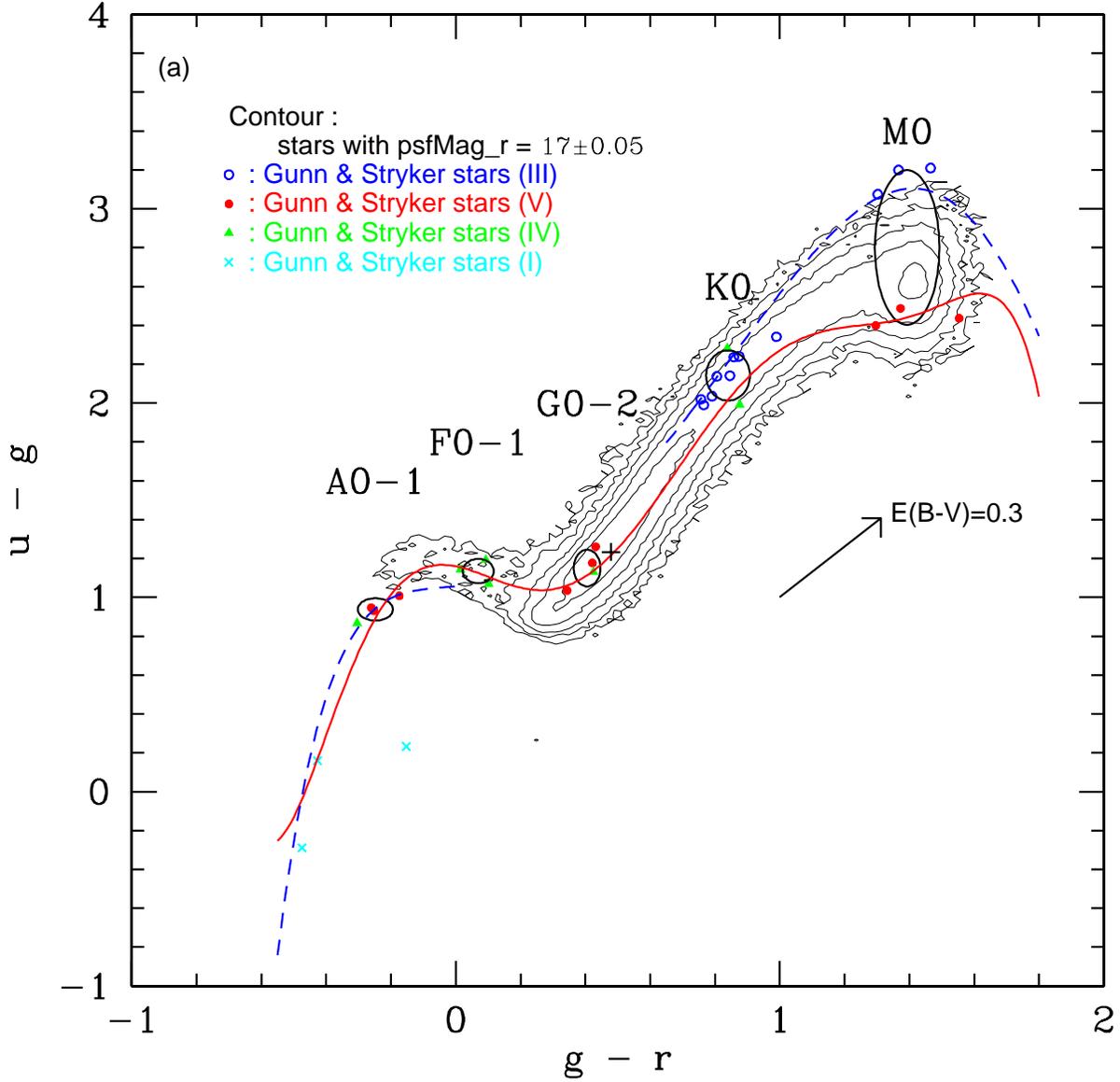}
\caption{(a) Distribution of stars in the SDSS general catalogue 
in $u-g$ vs. $g-r$ plane (the same as
Figure 2 but with isolated points representing individual stars suppressed).
Two curves are empirical loci of main sequence
stars obtained by interpolating the GS83 stars with known
spectroscopic
types separately for luminosity class V (solid curve, red) and 
III (dashed curve, blue). 
In panel (a) ellipses show the position of the GS83 stars
with types given as indicated in the figure with the size representing the
variance of the subsample. The plus symbol is colour of the Sun
given in the text. 
In panel (b) [ see next page ] line segments
indicate variations due to metallicity from [Fe/H]=$-1$ to +1
with 0.5 unit steps (blobs),
from downstairs up, 
using the Kurucz atmosphere model. Asterisks show the position for
[Fe/H]=$0$.
\label{fig:8}}
\end{figure}

We repeat in Figure \ref{fig:8}, the $u-g$ vs. $g-r$ colour-colour
distribution shown in Figure \ref{fig:2}. In panel (a) the loci of GS83
stars are represented by the two curves, the solid curve for 
luminosity class V and the dashed curve for class III.  These
curves are drawn by spline-interpolating the mean positions of the GS83
stars that are classified as III and V.  The curve segments that are not
constrained well by the data are skipped in the figure. The ellipses
are the areas where GS83 stars with some specified types, A0 and A1 stars
(denoted as A0$-$1),
F0 and F1 stars, G0, G1 and G2 stars, K0 stars and M0 stars fall 
(as indicated), where III and V are not
distinguished.  The size of the ellipses corresponds to the sample
variance.  The two curves cross at around $g-r=0$, but
the data are scanty for class III and this crossover is yet to be
examined with a larger data set. We see that the gap between the two
curves widens for $g-r>1$, which is likely to account for the gap between
the two branches
in the GS83 sample as we saw earlier, and as also noted by \citet{Yanny2009}. 
We indicate at the bottom right of the panels how reddening
affects colour. The length of the arrow corresponds to $E(B-V)=0.3$,
and the extinction curve we used \citep{ODonnell}
gives $k_\lambda=A_\lambda/E(B-V)$:
$k_u=5.155$, $k_g=3.793$ and $k_r=2.751$.

We give in Table \ref{tab:star} the effective temperature and colours
for given spectral types for stars in the GS83 sample, where the use
is made of our eq. (1) to estimate the effective temperature or, where 
available, the
temperature is adopted from the IRFM measurements \citep{DiBenedetto1998}.  We
calculate the mean over the available stars, although for most
listings only one or two stars are available in the GS83 sample.

It is expected that colour is affected by metallicity.
To show the range in $u-g$ vs. $g-r$ colour space that is covered by
the variation of metallicity, we indicate it in Figure \ref{fig:8}(b)
with line segments where the two edges correspond to [Fe/H]=+0.5
(upper edge) and $-$1 (lower edge) for luminosity class V, taking the
atmosphere model of \citet{CK2004}\footnote{\citet{CK2004} gives
   the prediction significantly different from \cite{Kurucz1993} for
   red stars with $g-r>1$ in this figure due to the inclusion of H$_2$O
   opacities and the revision of TiO lines.}.  The asterisk on the
segment shows the position of [Fe/H]=+0.  These line segments are
shifted for luminosity class III stars, somewhat irregularly for K and
M stars.  We expect that the 
Kurucz atmosphere models may give the relative metallicity-dependent 
position with reasonable accuracy, though the absolute 
position may somewhat be
shifted. The models properly
represent the range covered by the variation of metallicity.  This
shows that the width of the distribution of stars in the general
catalogue is consistent with the variation covered by the range of
metallicity $-1<$[Fe/H]$<0$ along with further widening of the range
due to the mixture of luminosity class III and V stars.

\begin{figure}
%\plottwo{figure9a.ps}{figure9b.ps}
\plotone{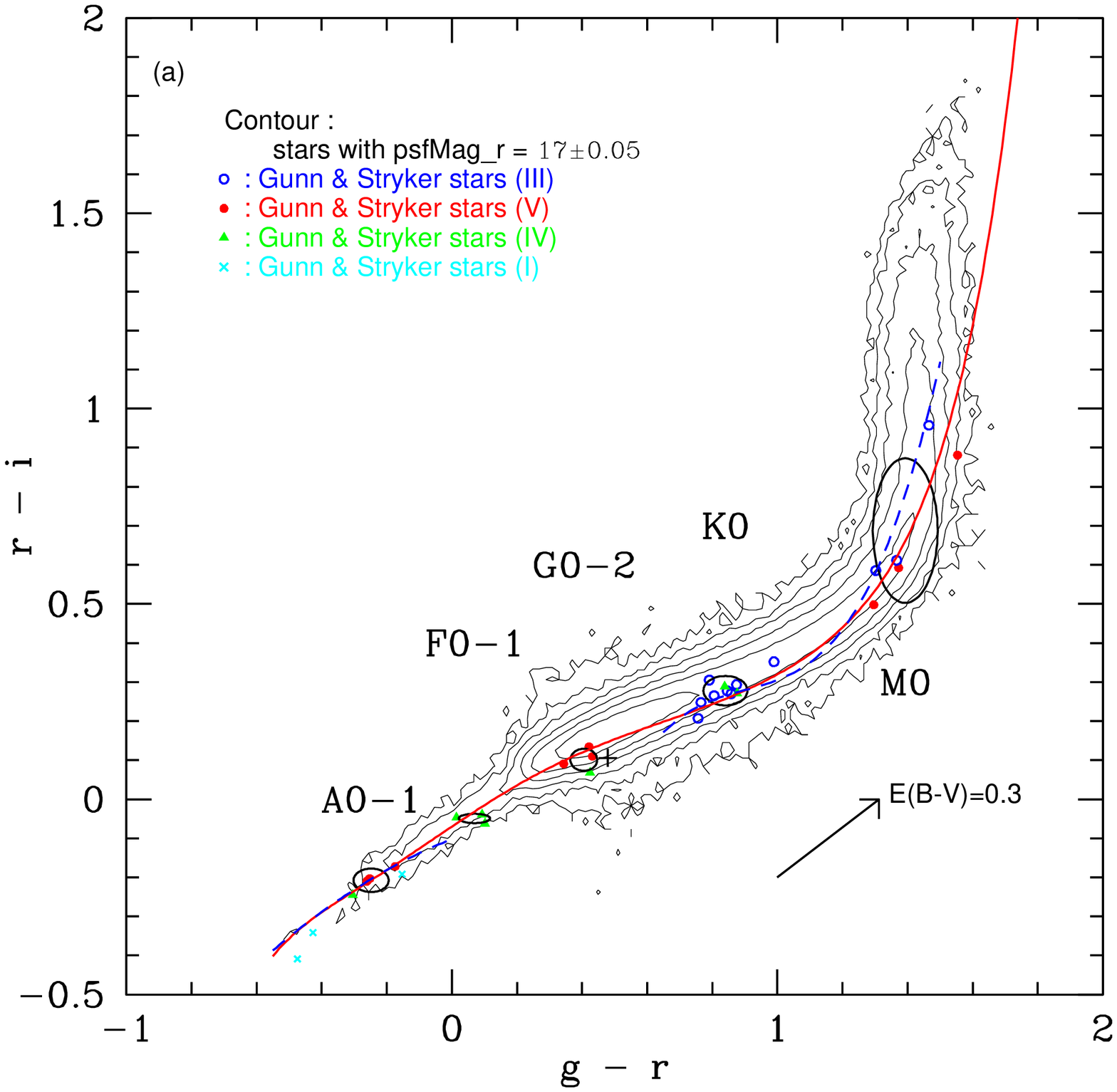}
\caption{(a) Same as Figure \ref{fig:8} but in $r-i$ vs. $g-r$ plane.
\label{fig:9}}
\end{figure}

A similar figure is given for the $r-i$ vs. $g-r$ colour distribution
in Figure \ref{fig:9}. The splitting of tracks between
luminosity classes III and V is very small: the two curves are nearly
degenerate except for red stars with $g-r>1.2$. 
The variation due to metallicity is also reduced although it is still 
substantial in $g-r$.
This generally smaller variation makes the discrepancy between the
observation and the Kurucz atmosphere model, indicated by line
segments in Figure \ref{fig:9}b, somewhat more apparent. The Kurucz atmosphere
model fits the curve representing the data only at its lowest
metallicity edge, although the discrepancy is only about 0.03 mag in
$r-i$.
The trend is similar for the $i-z$ vs. $g-r$ colour
distribution (Figure \ref{fig:10}). We see a small gap appearing between the
two luminosity classes which is somewhat more conspicuous than is seen in $r-i$
vs. $g-r$. Metallicity induced colour variation is larger for $i-z$,
even larger than for $u-g$, for
red stars.

All of the above indicates that the SDSS system is relatively close
to a true AB system; the suspected corrections to an AB system of a
few hundredths of a magnitude \citep{DR7} are small enough not to have
been seen in this analysis.

\section{Colours of other objects}

To give the idea as to the feature of SDSS colours. we show in Figure
\ref{fig:11} the distribution of white dwarfs, the sample of which is
taken from the 9300 spectroscopically confirmed white dwarfs
($16.75<r<17.25$) of \citet{Eisenstein2006}.  They are white dwarfs
with temperature typically $>7000$K. Cooler white dwarfs are
degenerate with main sequence stars, and more with
subdwarfs. Selections other than colours are needed to find candidates
for cooler white dwarfs \citep{kilic06, harris06}. The locus of
quasars is also added (solid curve) taking the composite spectrum of
\citet{VandenBerk2001} as the fiducial and by redshifting it to $z\leq
2.5$. The lowest redshifts shown depend on colour: they are 0 for
$u-g$, 0.05 for $g-r$, and 0.35 for $r-i$. The locus of Type Ia
supernovae at the epoch of their B-band brightness maximum is also
shown (dashed curve) using the model spectrum of SALT2 \citep{SALT2}
that is moved to non-zero redshift, $z\leq 0.5$.  Some parts of the
locii lie outside the frame of the figure.

The figure shows the distribution of white dwarfs well separated from
normal stars due to their UV excess in the $u-g$ vs. $g-r$ plane, in
so far as hot white dwarfs (approximately with $T>7000$K) are
concerned, as efficiently used in \citet{Eisenstein2006}, and to some
extent also in the $i-z$ vs.  $g-r$ plane. A strong degeneracy with
normal stars is seen in the $r-i$ vs. $g-r$ plane. White dwarfs are
degenerate with the blue tip of the distribution of normal stars with
these colours.

Low redshift quasars, having $u-g$ colors much bluer than most stars,
are also well separated from normal stars on the $u-g$ vs. $g-r$ plane
till quasars reach $z\approx 2.5$, where they mingle into the stellar
locus.  The confusion of quasars with white dwarfs may take place in
the $u-g$ vs.  $g-r$ plane, though quasars are somewhat redder in
$g-r$: this degeneracy is lifted if the $i-z$ vs. $g-r$ colour
is used in addition.  This demonstrates the advantage of multicolour space
in the photometric selection of low redshift quasar candidates. The
photometric target selection for quasars is discussed at length in the
papers of \citet{Fan2001} and \citet{Richards2002, Richards2004}, and
we do not discuss this problem further.

The figure shows that Type Ia supernovae can be identified, well separated
from the stellar locus,
given the homogeneity of
colours of Type Ia supernovae. The fiducial colour of the SALT II template
at zero redshift, $u-g=0.476$, $g-r=-0.170$ and $r-i=-0.673$, is
compared with that derived from the SDSS supernova sample, which reads
$u-g=0.591\pm 0.237$, $g-r=-0.093\pm 0.157$ and $r-i=-0.636\pm0.088$
where the error represents the sample variance \citep{Yasuda2010}.
The diagram shows that Type Ia supernova locus moves
monotonically with redshift, indicating that the photometric redshift can
be estimated with a good confidence, given good colors at maximum light.

\begin{figure}
\plottwo{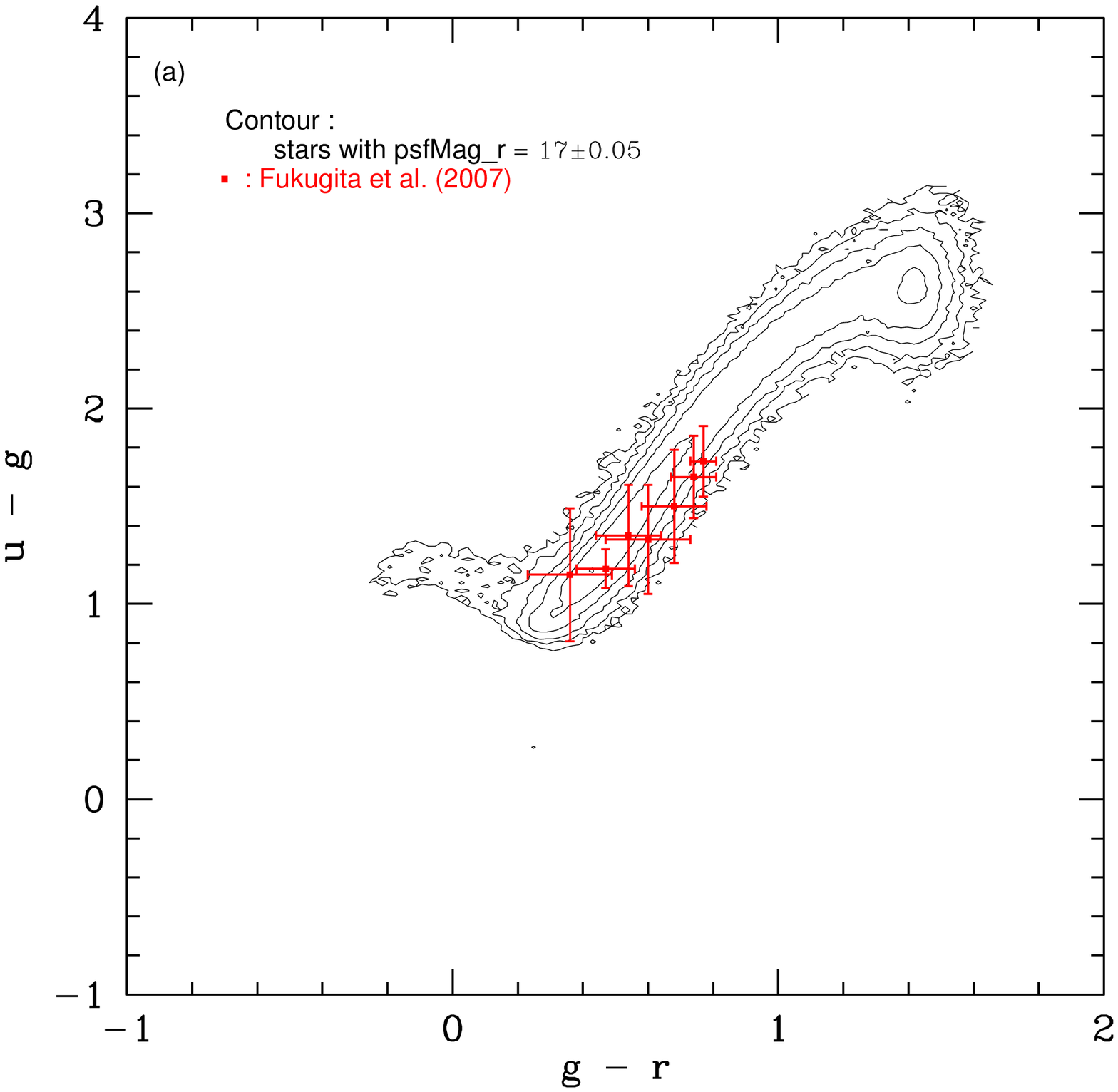}{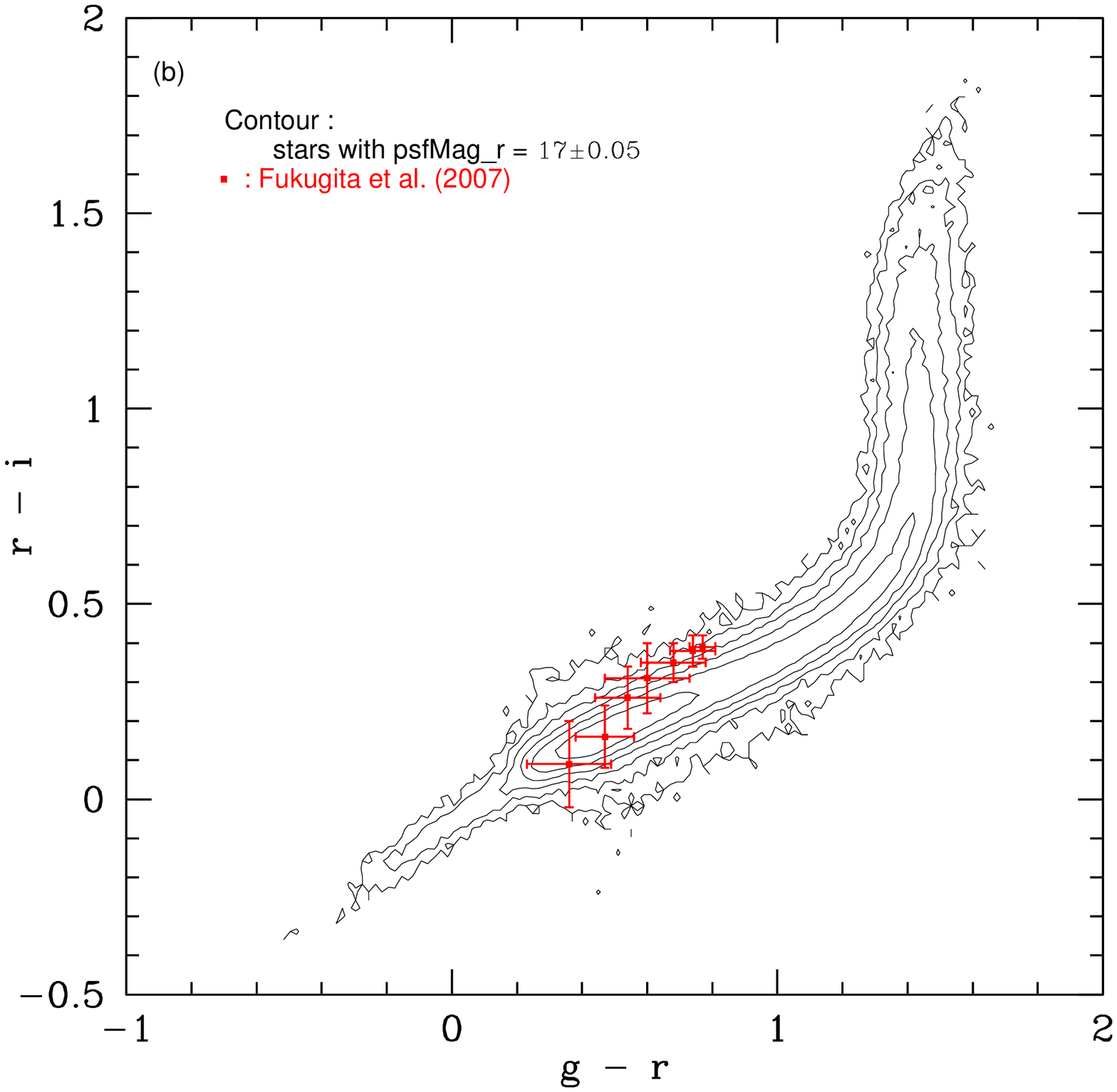}
\epsscale{0.5}
\plotone{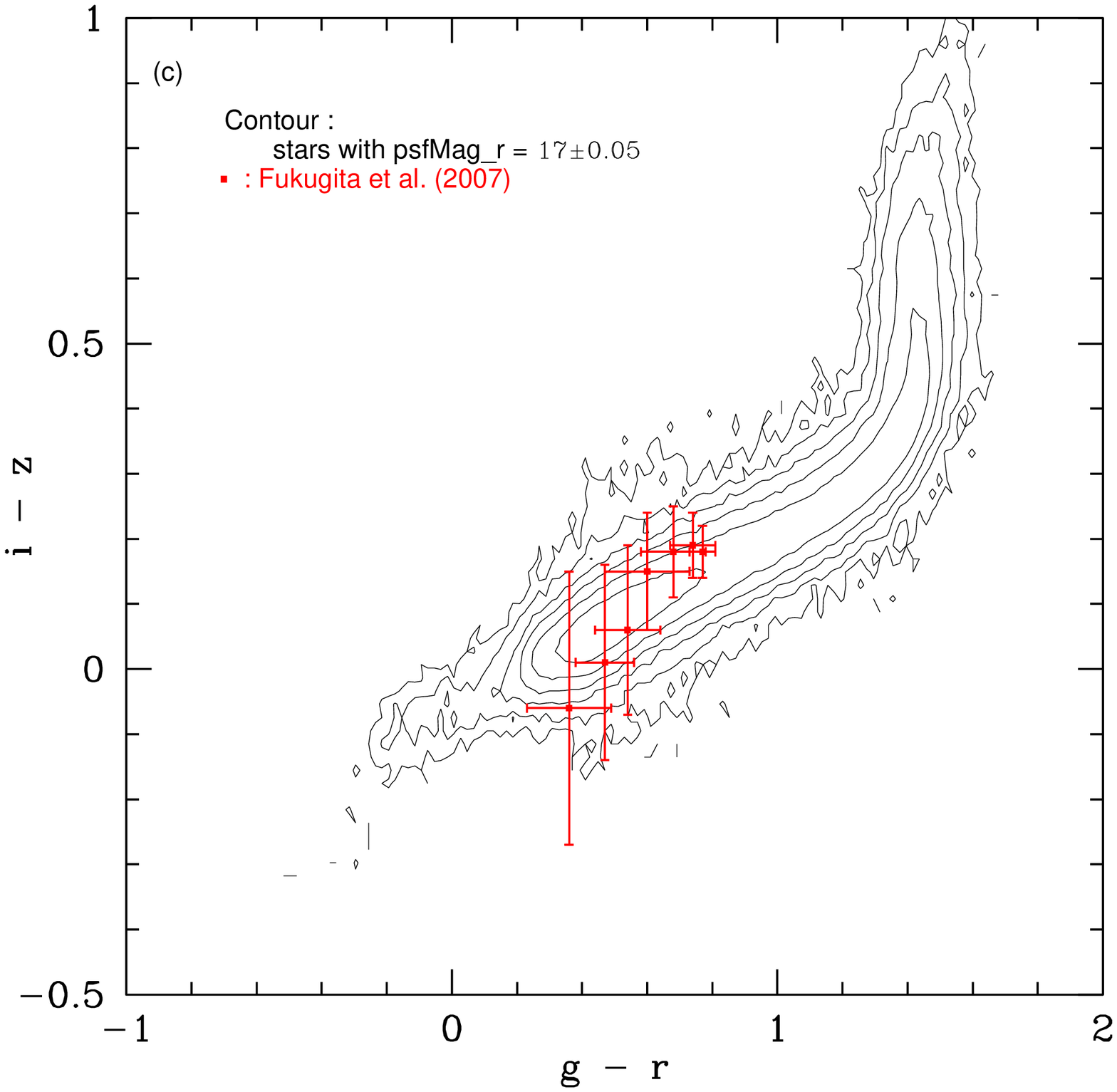}
\epsscale{1.0}
\caption{Distribution of stars in (a) $u-g$ vs. $g-r$ plane, (b)
$r-i$ vs.  $g-r$ plane, and (c) $i-z$ vs.  $g-r$ plane, where colours of
galaxies are overlaid. The error bars represent the variance of colours
in the morphologically classified sample, from the right to the left
E, S0, Sa, Sb, Sc, Sd and Im.
\label{fig:12}}
\end{figure}

The three panels of Figure \ref{fig:12} show colours of
morphologically classified galaxies at $z\approx 0$. Galaxy colours
are taken from \citet{Fukugita2007} based on the visual classification
of 2250 galaxies with $r\leq 16$ in the northern equatorial
stripe. Magnitudes are Petrosian.  The error bars indicate the
variance of the samples for E, S0, Sa, Sb, Sc and Im.  Galaxies occupy
a colour space locus narrower than stars, with colours roughly
corresponding to F5$-$K2 stars in the range $g-r=0.2-0.8$.

\section{Brightness and colour of the Sun}

The brightness of the Sun is taken as a basic unit in
astrophysical work. The measurement of the brightness of the Sun, however,
is notoriously difficult due to its extreme brightness and angular extent
\citep[e.g.,][]{Hayes1985}.  The measured broad band colour
of the Sun has been variously reported from $(B-V)_\odot=0.62$ to 0.68
by various investigators.
The choice
often taken is $(B-V)_\odot=0.65$ \citep{Allen1973}.

We adopt the absolute spectrophotometry of the Sun measured by SOLSPEC
in the {\it Atmospheric Laboratory for Application and Science}
(ATLAS) 3 mission \citep{Thuillier2003}, the absolute flux of which
has been calibrated against a black body standard.  We
present in Table \ref{tab:solar} the synthetic broad band magnitudes
of the Sun for $V$, $B-V$, $g$, $u-g$, $g-r$, $r-i$ and $i-z$. We take
the $B$ and $V$ response functions evaluated by \citet{AS1969}, which
seem to give the minimal offset of synthetic brightness against
Johnson broad band photometry among a number of response functions we
tested \citep{FSI1995}.  We also present synthetic values of
brightness of the Sun using another spectrophotometric table compiled
by \citet{Colina1996} in order to indicate the size of possible
systematic errors.

The agreement among the different spectral energy distribution (SED)
is within about 0.02 mag for the V band brightness.  The SED of
\citet{Thuillier2003} integrated with the reference response function
of the SDSS 2.5m telescope with 1.3 airmasses, gives $V=-26.73$ mag
which agrees well with $-26.75\pm0.06$ mag of the summary value of
photometry, given by \citet{Hayes1985} and is also close to the
value, $-26.74$, adopted in Allen's {\it Astrophysical Quantities}
(1973).  The use of the SED from earlier
missions of ATLAS also gives V band brightness within 0.02 mag of the
ATLAS3 value. This is, in fact, the order of magnitude
\citet{Thuillier2003} claimed as the error of calibration.

This result leads us to expect that solar SED can also be used to
calculate broad band colours.  The $B-V$ colour from
\citet{Thuillier2003} is $0.62\pm0.01$.  While this looks somewhat
bluer than the conventional value, we should remark that $B-V$ colour
estimated from the colour temperature relations for main sequence
stars with the solar abundance also lie bluer, in the range 0.616$-$0.635
\citep[see][]{SF2000}.  The colour often quoted, $B-V=0.650$, is
rather close to colour of G4 stars than G2, and is suspected to be too
red.

The modern SED of the Sun gives a reasonable agreement among the
different data sets for both absolute flux and colour.  We
calculate with a reasonable confidence $g-r$, $u-g$, $r-i$ and $i-z$
colour of the Sun as given in Table \ref{tab:solar} in the AB$_{95}$
magnitude system. The $u-g$ and $g-r$ colours thus obtained (see
the cross symbols in Figure \ref{fig:8}) fall close
to the later-type edge of the 1 $\sigma$ ellipse representing G0$-$G2
stars of the GS83 sample in Figure \ref{fig:8}: $(u-g)_\odot=1.21$ and
$(g-r)_\odot=0.45$ are close to the mean of two G2V stars in the
GS83 sample, 1.22 and 0.43, respectively.  The same is true also
for the redder passbands:
$(r-i)_\odot=0.13$ and $(i-z)_\odot=0.02$ are compared with 0.12 and
0.03 of the GS83 stars.  The Sun thus matches very well with G2 stars.
The brightness in the $g$ passband is $-26.46$. Our $V$ and $g$
magnitudes translate to absolute brightness $M_V(\odot)=4.84$ and
$M_g(\odot)=5.11$ taking the distance modulus $-$31.57.  For other
colours, see Table 2.

\section{Summary}

We have studied photometric properties of stars given in the data
archive of the SDSS, the primary question being whether the
photometric calibration was, overall, properly done. 
We found that, over the entire lifetime of
the survey, the photometric calibration for point sources has been
made tightly against the SDSS standard stars and colours of the stars
are well defined.  We have not identified a sample of stars which
are significantly and systematically deviant from the SDSS standard stars.
%or stars that were spectro studied with a high accuracy.
%we do not observe any signatures that indicate a
%shift of colours of stars in the general catalogue of SDSS against the
%standard star sample, which is supposed to be tied to the AB$_{95}$ system. 
We have also found that the synthesised colour from the Gunn
Stryker spectrophotometric sample represents very well the colour of
stars in the SDSS general catalogue and vice versa.  Photometric
properties are perfectly consistent mutually with each other among the
three samples we studied.  It is also gratifying that the the synthesis of the
SDSS spectrophotometric data with the use of the reference response
function of the 2.5m telescope of the SDSS give broad band fluxes that
agree with broad band photometry of SDSS with little tilt along
colours.

We have also given the fiducial colours and temperatures for stars
empirically given spectral types, and show how metallicity and
surface gravity affect colours.  This enables us to infer spectral
type of stars when SDSS colours is given.
We show that $g-r$ colour can be used
as a good estimator of the effective temperature of stars. The
distribution of stars in colour space matches well with what we expect
from the variations of metallicity and surface gravity. We also
present the brightness and colour of the Sun with synthetic calculations,
which shows that colour of the Sun matches well with G2V stars in the
SDSS photometric system.

A problem we encountered in the present study is that the sample of
stars with well known temperature and metallicity is still too scanty, and 
bright stars
that have  SDSS photometry are small in number. This is
especially depressing when we require these two quantities at the same time.
There is a pressing need to acquire photometry with the SDSS passbands
for bright star samples which have good measured atmospheric parameters,
(composition, temperature, and gravity) 
to connect more tightly SDSS photometry with physical
stellar parameters.

\vskip10mm\noindent
{\bf Acknowledgement}

We are grateful to Maki Sekiguchi and Takashi Ichikawa for their
collaboration in our earlier work to define the AB system using the solar
spectroscopic data (Sect. 5), and Masayuki Tanaka for his work measuring
the response functions of the SDSS imager.  We also thank David Hogg and
Heather Morrison for a number of comments that helped us to tighten
our results. 
MF is supported by the Monell
Foundation at the Institute in Princeton, and received a Grant in Aid of
the Ministry of Education (Japan) in Tokyo.

Funding for the SDSS and SDSS-II has been provided by the Alfred P.
Sloan Foundation, the Participating Institutions, the National Science
Foundation, the U.S. Department of Energy, the National Aeronautics
and Space Administration, the Japanese Monbukagakusho, the Max Planck
Society, and the Higher Education Funding Council for England. 
%The SDSS Web Site is http://www.sdss.org/.
The SDSS is managed by the Astrophysical Research Consortium for the
Participating Institutions. The Participating Institutions are the
American Museum of Natural History, Astrophysical Institute Potsdam,
University of Basel, University of Cambridge, Case Western Reserve
University, University of Chicago, Drexel University, Fermilab, the
Institute for Advanced Study, the Japan Participation Group, Johns
Hopkins University, the Joint Institute for Nuclear Astrophysics, the
Kavli Institute for Particle Astrophysics and Cosmology, the Korean
Scientist Group, the Chinese Academy of Sciences (LAMOST), Los Alamos
National Laboratory, the Max-Planck-Institute for Astronomy (MPIA),
the Max-Planck-Institute for Astrophysics (MPA), New Mexico State
University, Ohio State University, University of Pittsburgh,
University of Portsmouth, Princeton University, the United States
Naval Observatory, and the University of Washington.

%--------------------------------------------------------------------------------
%--------------------------------------------------------------------------------
%\section{bibliography}

\newpage
\begin{deluxetable}{crrrrrcrrrrr}
\tablecolumns{10}
\tablewidth{0pc}
\tablecaption{Temperature and colour of stars.
\label{tab:star}} 
\tablehead{
\colhead{Type} & \colhead{$T_{\rm eff}$} & \colhead{$T_{\rm eff}$} & \colhead{$u-g$} & \colhead{$g-r$} & \colhead{$r-i$} &\colhead{Type} & \colhead{$T_{\rm eff}$} & \colhead{$T_{\rm eff}$} & \colhead{$u-g$} &\colhead{$g-r$} & \colhead{$r-i$}\\
\colhead{} & \colhead{colour} & \colhead{IRFM} & \colhead{} & \colhead{} &
\colhead{} & \colhead{} & \colhead{colour} & \colhead{IRFM} & \colhead{} &\colhead{}}
\startdata
B2V  & (11195) &      &  $-0.08$ & $-0.50$ & $-0.37$ & B2III & (11180) &      & $-0.10$ & $-0.50$ & $-0.35$ \\
B7V  &  (9631) &      &  $0.49$  & $-0.34$ & $-0.25$ & B7III &  (9832) &      & $0.49$ & $-0.36$ & $-0.26$ \\
A0V  &  8715 &      &    $0.98$ & $-0.22$ & $-0.19$ & A0III &  ---  &      &        & \\
A5V  &  7656 &      &    $1.08$ & $-0.04$ & $-0.09$ & A5III &  7812 & 7997 &   $1.10$ &$-0.07$ & $-0,13$ \\
F0V  &  ---  &      &           &     &           & F0III &  ---  &      &  &       \\
F4V  &  6831 &      &    $1.11$ & $0.13$ & $-0.11$ & F4III &  ---  &      & &        \\
F9V  &  5912 & 5902 &    $0.90$ & $0.37$ & $0.10$ & F9III &  ---  &      & &        \\
G0V  &  6008 &      &    $1.04$ & $0.34$ & $0.09$ & G0III &       &      &         \\
G2V  &  5747 &      &    $1.22$ & $0.43$ & $0.12$ & G2III &       &      & &        \\
G5V  &  5596 &      &    $1.29$ & $0.48$ & $0.14$ & G5III &       &      &         \\
G8V  &  5217 & 5155 &    $1.51$ & $0.62$ & $0.21$ & G8III &  4969 & 5026 & $1.86$ &  $0.72$ & 0.23 \\
K0V  &  5610 &      &    $1.19$ & $0.47$ & $0.15$ & K0III &  4731 &      & $2.14$ &  $0.84$ & 0.28 \\
K4V  &  4600 &      &    $2.06$ & $0.90$ & $0.29$ & K4III &  4017 & 4050 & $3.09$ &   $1.24$ & $0.43$ \\
K7V  &  3879 & 4011 &    $2.47$ & $1.34$ & $0.54$  & K7III &                                \\
M0V  &  (3793) &      &  $2.44$ & $1.41$ & $0.66$ & M0III &  3828 & 3846 & $3.16$ &   1.38  & 0.72\\
M2V  &  (3700) &      &  $2.47$ &  $1.48$ & $0.92$ & M2III &  (3585) &      &   3.10 & 1.57  & 1.28\\
M5V  &  (3415) &      &  $2.29$ &  $1.72$ & $2.12$ & M5III &  (3576) &      &   2.60 & 1.58  & 1.89 \\
\enddata
\tablecomments{$T_{\rm eff}({\rm colour})$ is estimated from eq.(1)
and the applicability is suspected for values given in parentheses; see the
text.}
\end{deluxetable}

\begin{deluxetable}{lrrrrrrr}
\tablecolumns{8}
\tablewidth{0pc}
\tablecaption{Brightness and colour of the Sun.
\label{tab:solar}} 
\tablehead{
\colhead{SED Data} & \colhead{$V$} & \colhead{$B-V$} & \colhead{$g$} & \colhead{$u-g$} & \colhead{$g-r$} & \colhead{$r-i$} & \colhead{$i-z$}}
\startdata
\citet{Thuillier2003} & $-26.734$ & 0.622 & $-26.459$ & 1.213 & 0.451 & 0.127 & 0.015 \\
%Thuillier et al. & -26.749 & 0.622 & 1.239  & 0.465   & 0.128 &  0.016  \\
%Solspec          & -26.749 & 0.622 & ....   & 0.465   & 0.129  \\
\citet{Colina1996}    & $-26.748$ & 0.644 & $-26.453$ & 1.234 & 0.479 & 0.106 & 0.020  \\
\tableline
\tableline
  & \colhead{$M_V$} & & \colhead{$M_u$} & \colhead{$M_g$} & \colhead{$M_r$} & \colhead{$M_i$} & \colhead{$M_z$} \\
\tableline
  & 4.84 & & 6.32 & 5.11 & 4.66 & 4.53 & 4.52 \\
\enddata
\tablecomments{Synthetic broad band absolute brightness
obtained using Thuillier et al. (2003)'s SED.
$V$ magnitude is in the Johnson system, giving
$V=0.03$ for $\alpha$ Lyr.  $ugriz$ magnitudes are in the
$AB_{95}$ system.}
\end{deluxetable}

\begin{figure}
\plotone{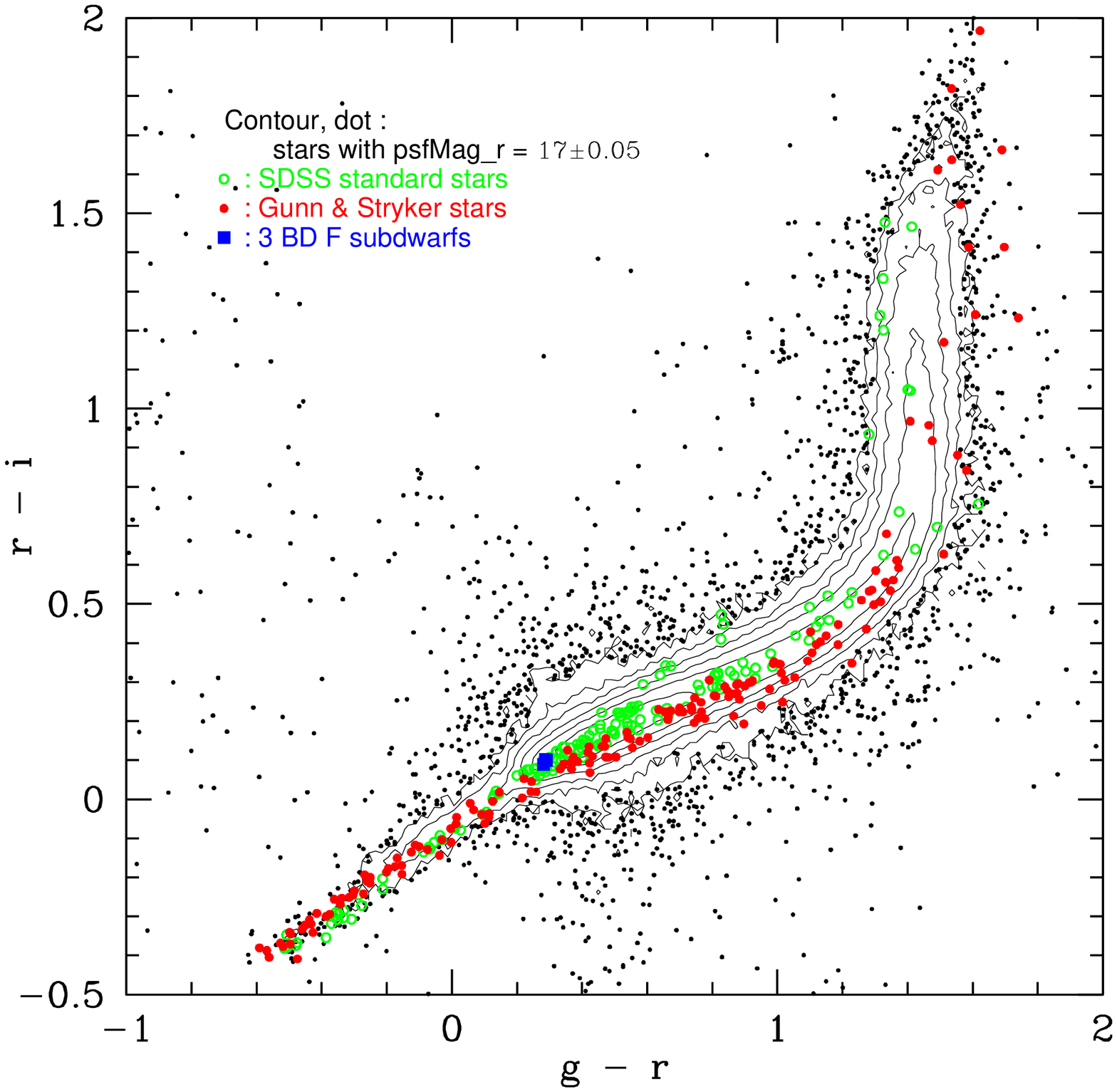}
\caption{Same as Figure 2, but for $r-i$ vs. $g-r$.
\label{fig:5}}
\end{figure}

\begin{figure}
\plotone{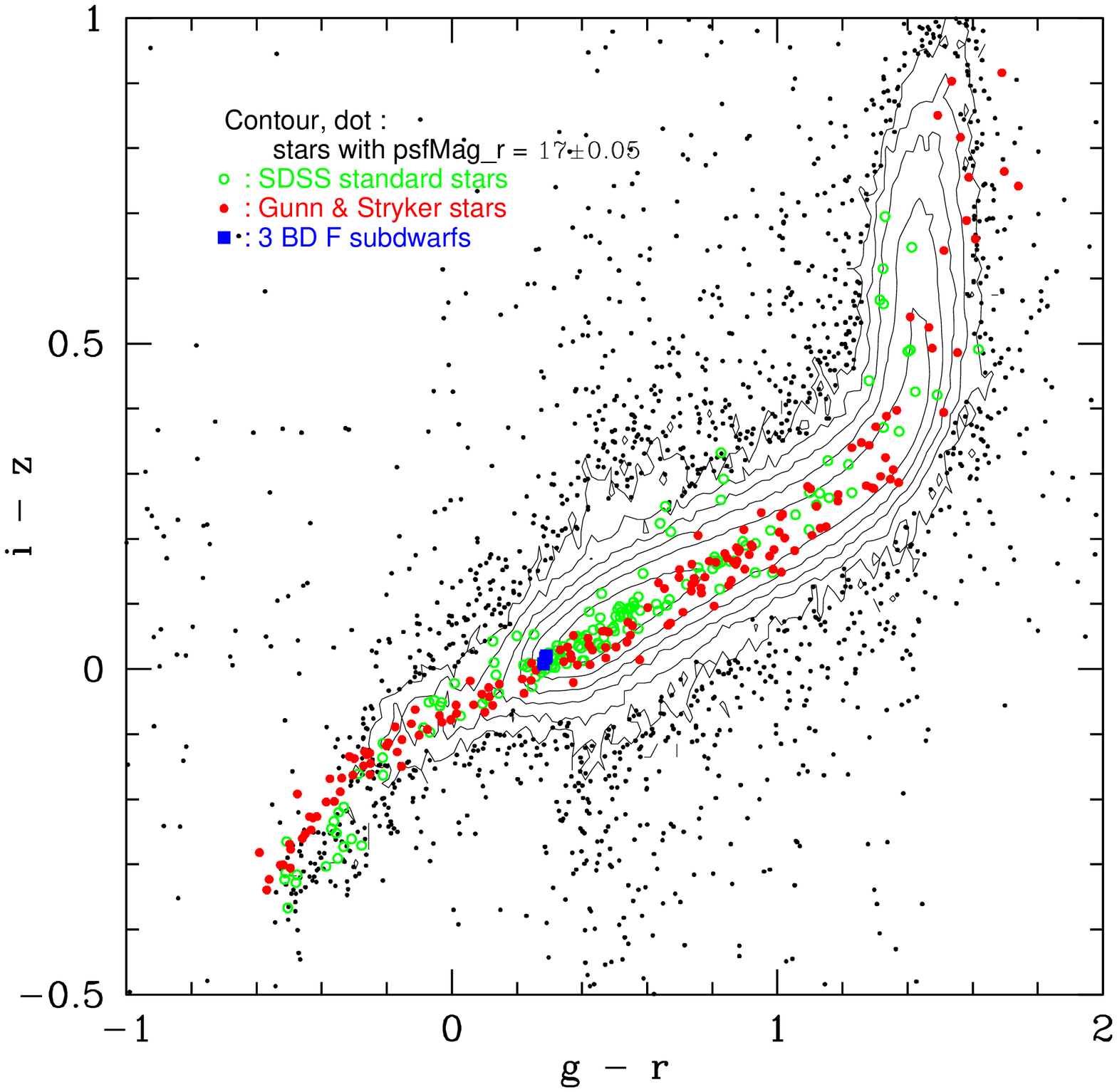}
\caption{Same as Figure 2, but for $i-z$ vs. $g-r$.
\label{fig:6}}
\end{figure}

\begin{figure}
\figurenum{8}
\plotone{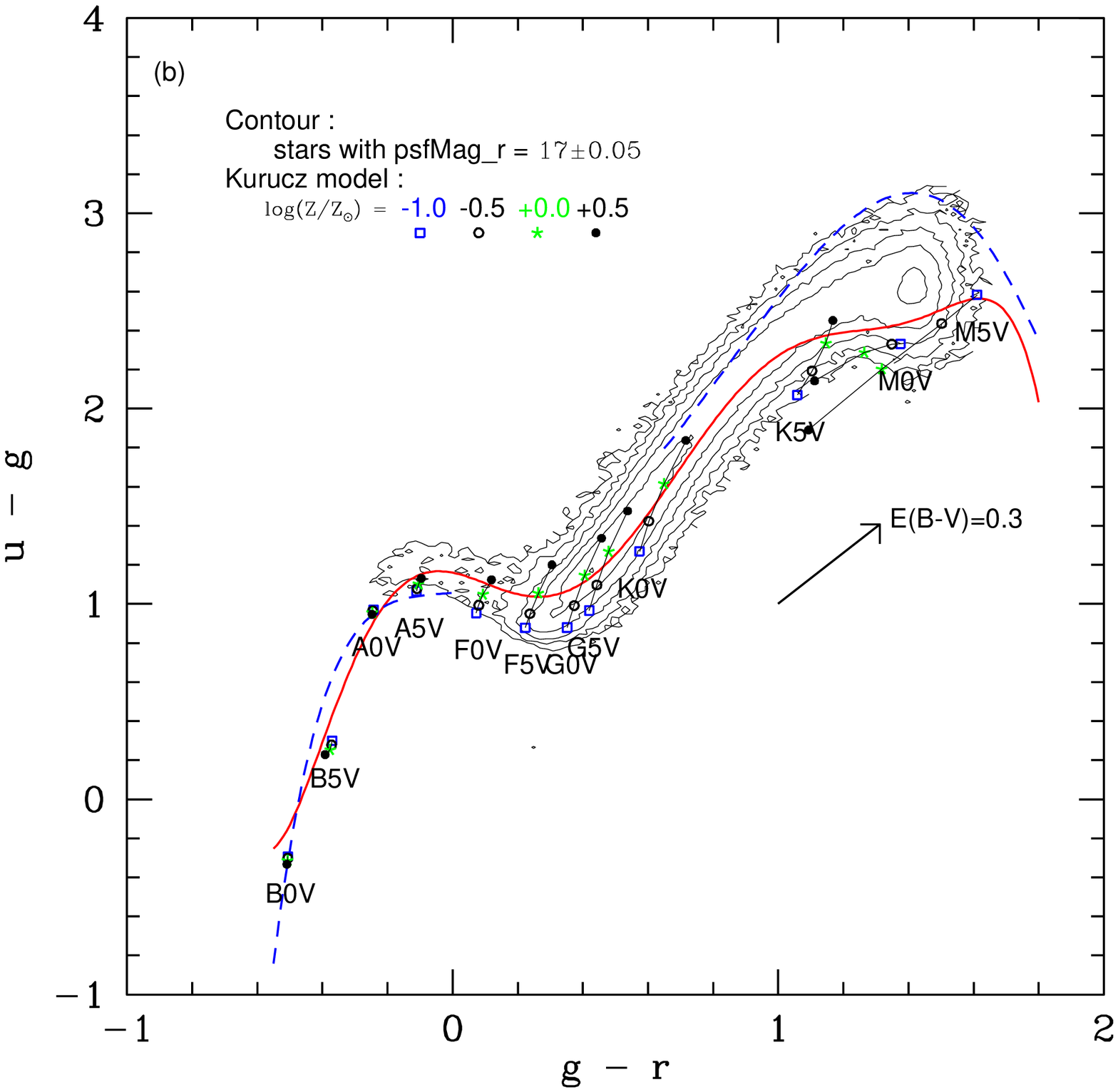}
\caption{(b) continued}
\end{figure}

\begin{figure}
\figurenum{9}
\plotone{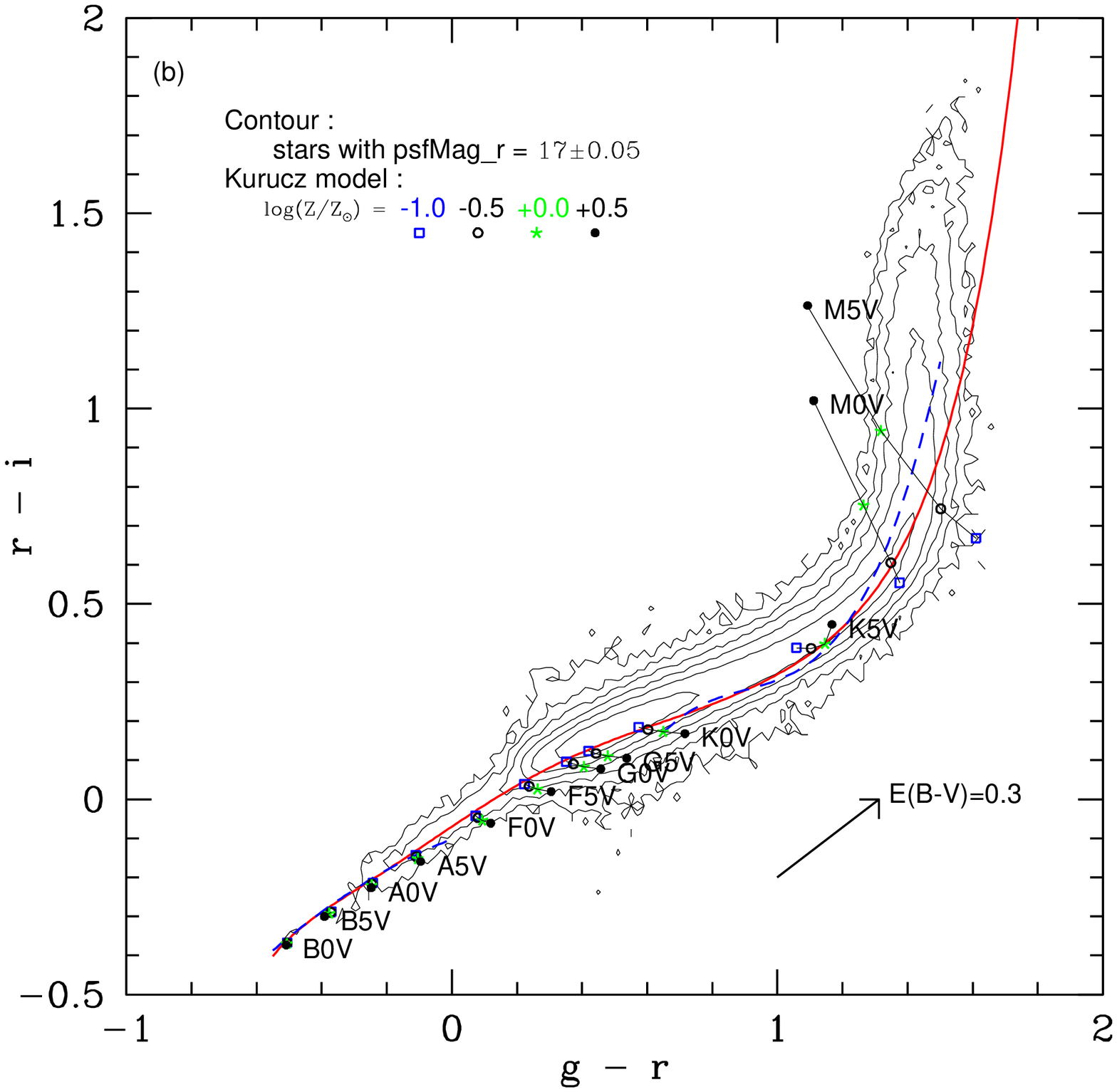}
\caption{(b) continued}
\end{figure}

\begin{figure}
%\plottwo{figure10a.ps}{figure10b.ps}
\plotone{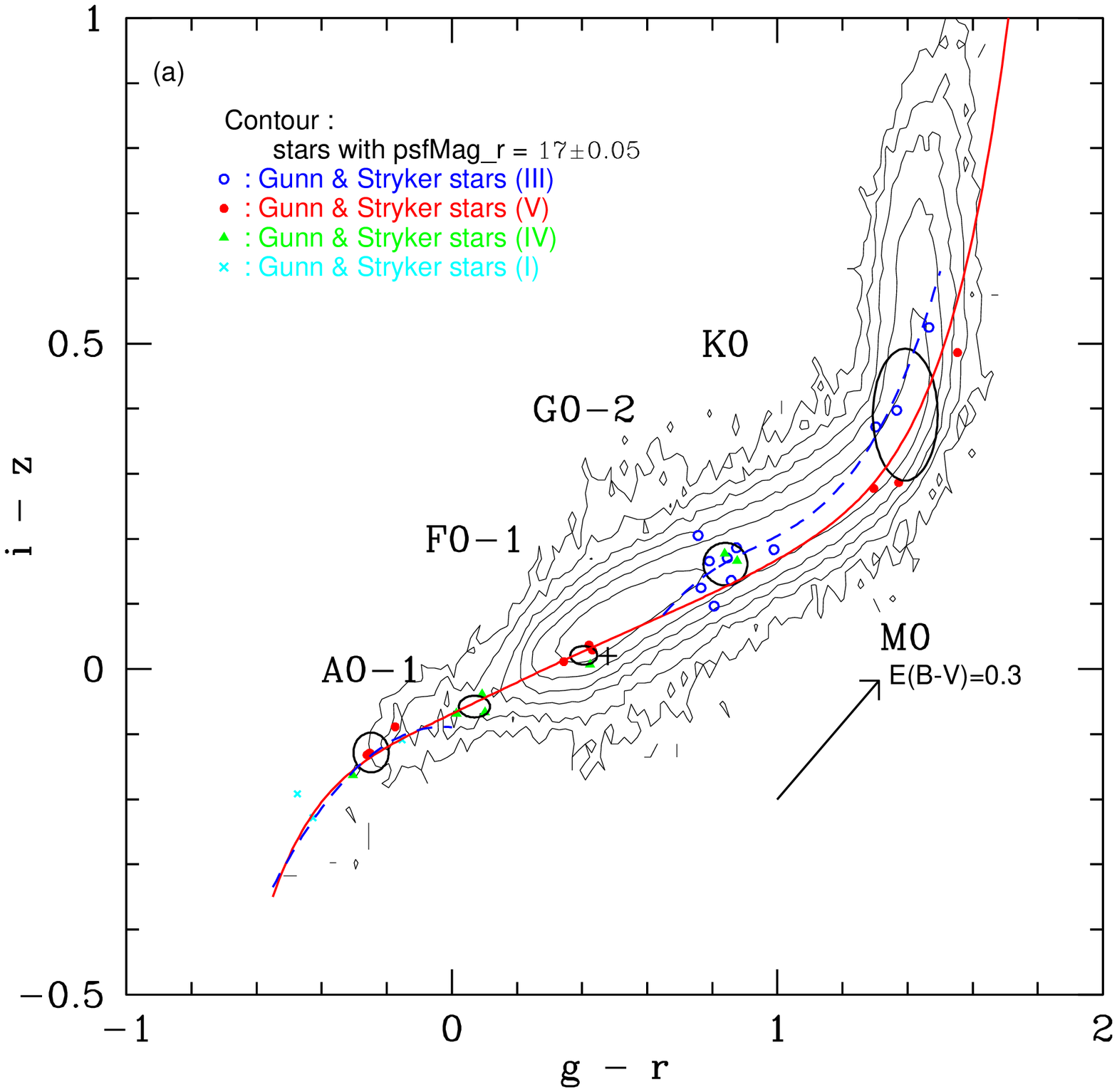}
\caption{(a) Same as Figure \ref{fig:8} but in $i-z$ vs. $g-r$ plane.
\label{fig:10}}
\end{figure}

\begin{figure}
\figurenum{10}
\plotone{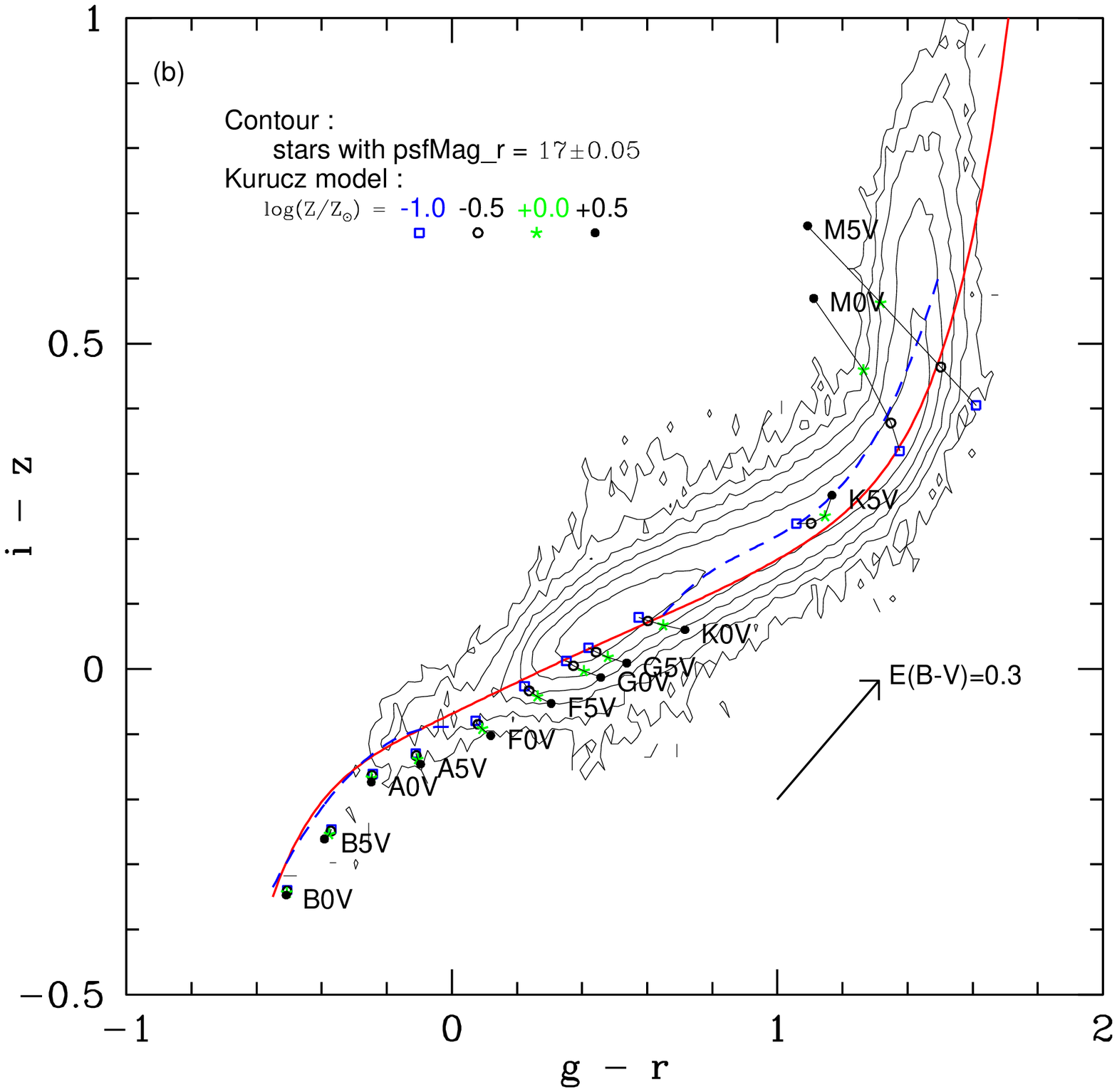}
\caption{(b) continued}
\end{figure}

\begin{figure}
\plottwo{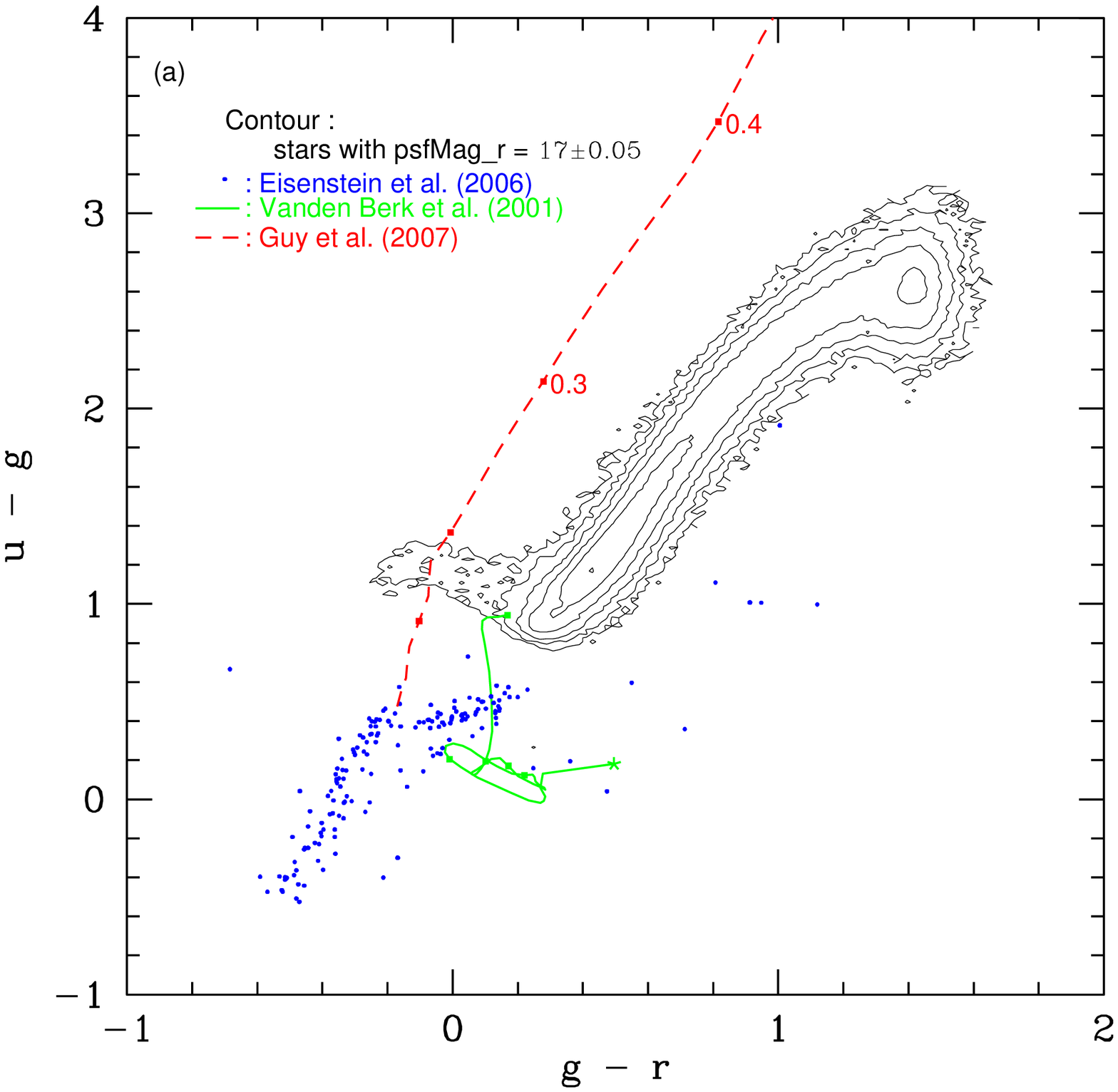}{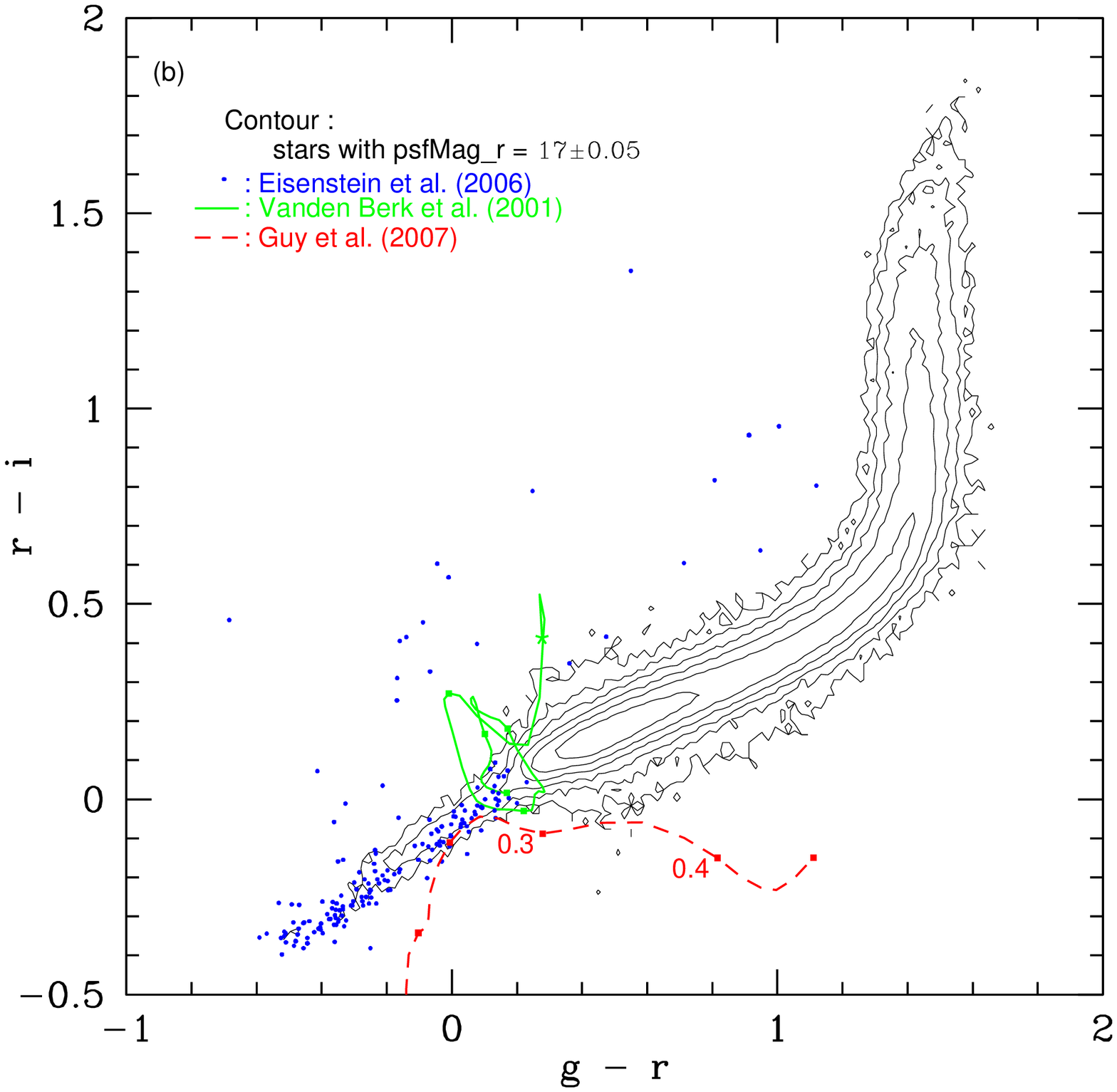}
\epsscale{0.5}
\plotone{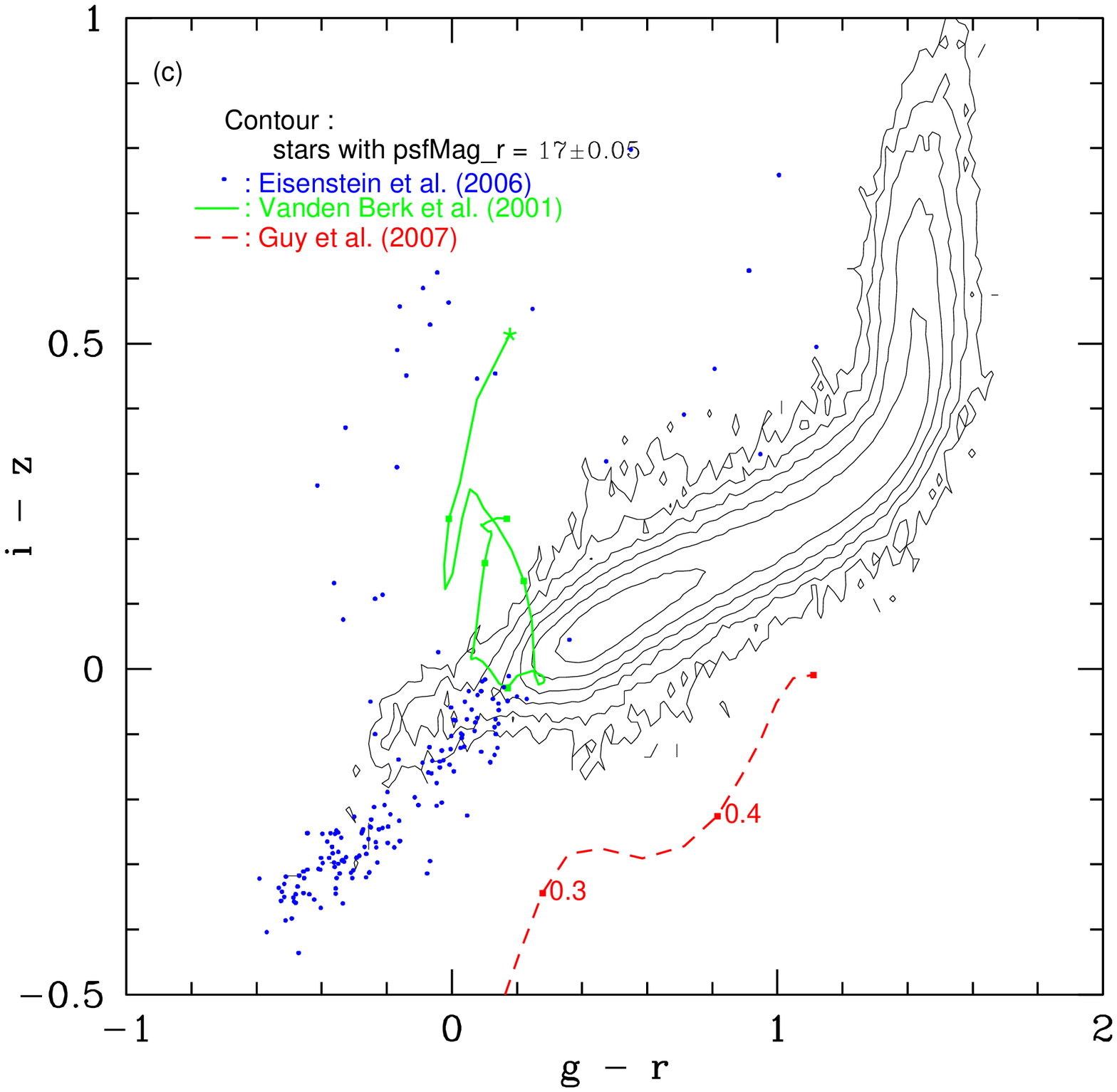}
\epsscale{1.0}
\caption{Distribution of stars in (a) $u-g$ vs. $g-r$ plane, (b) $r-i$
   vs.  $g-r$ plane, and (c) $i-z$ vs.  $g-r$ plane (the same as Figure
   \ref{fig:2}, \ref{fig:5} and \ref{fig:6}, respectively, with
   isolated points representing individual stars suppressed).  Small
   dots show hot white dwarfs. The solid curve shows the locus of quasars
   having the composite spectrum with redshift $\approx 0$ to 2.5, with blobs at
   the 0.5 interval in $z$: the lowest redshift is shown
   with the asterisk, which is $z=0$ for (a), $z=0.05$ for (b) and
   $z=0.35$ for (c).
   The dashed curve is the locus of Type Ia supernovae redshifted 0 to
   $z\leq 0.5$ towards red. The blobs on the supernova curve are at a
   0.1 interval in redshift,
   but those at $z\approx0$ is out of the range of the figure and
   omitted in (b) and (c). The blobs at $z=0.1$ and 0.2 are also omitted in (c).
   \label{fig:11}}
\end{figure}

\end{document}